\documentclass[%
reprint,
superscriptaddress,
 amsmath,amssymb,
 aps,
pra,
floatfix,
twocolumn 
]{revtex4-2}

\usepackage{graphicx}
\usepackage{dcolumn}
\usepackage{bm}
\usepackage{amsmath}
\usepackage[utf8]{inputenc}
\usepackage{physics}
\usepackage{array}
\usepackage{makecell}
\usepackage{multirow}
\usepackage{booktabs} 
\usepackage{tabularx}
\usepackage{mathrsfs}
\usepackage{xcolor}
\usepackage{quantikz}
\usepackage{color}
\usepackage{float}
\usepackage{appendix}
\usepackage{enumitem}
\usepackage{natbib,hyperref}
\usepackage{xr}
\newcommand{\quotes}[1]{``#1''}

\hypersetup{colorlinks=true,linkcolor=blue,urlcolor=blue,citecolor=blue}

\newcommand{\RN}[1]{%
  \textup{\uppercase\expandafter{\romannumeral#1}}%
}

\begin{document}

\preprint{APS/123-QED}

\title{Efficient Fermi-Hubbard model ground-state preparation by coupling\\to a classical reservoir in the instantaneous-response limit}

\author{Zekun He}
 \email{zh168@georgetown.edu}
 \affiliation{%
Department of Physics, Georgetown University, Washington DC 20057, USA
}%

\author{Lorenzo Del Re}
\affiliation{
Institute for Theoretical Physics and Astrophysics,
Julius-Maximilians-Universit\"at W\"urzburg,
Am Hubland,
97074 W\"urzburg,
Germany
}

 \author{A.~F.~Kemper}
\email{akemper@ncsu.edu}
\affiliation{Department of Physics, North Carolina State University, Raleigh, North Carolina 27695, USA}

\author{J. K. Freericks}%
\email{james.freericks@georgetown.edu}
\affiliation{%
Department of Physics, Georgetown University, Washington DC 20057, USA
}%

\date{\today}

\begin{abstract}
Preparing the ground state of the Fermi-Hubbard model is challenging, in part due to the exponentially large Hilbert space, which complicates efficiently finding a path from an initial state to the ground state using the variational principle. In this work, we propose an approach for ground state preparation of interacting models by involving a classical reservoir, simplified to the instantaneous-response limit, which can be described using a Hamiltonian formalism. The resulting time evolution operator consist of spin-adapted nearest-neighbor hopping and on-site interaction terms similar to those in the Hubbard model, without expanding the Hilbert space. We can engineer the coupling to rapidly drive the system from an initial product state to its interacting ground state by numerically minimizing the final state energy.
This ansatz also closely resembles the Hamiltonian variational ansatz, offering a fresh perspective on it.
\end{abstract}

\maketitle 

\section{Introduction} 

Ground-state preparation, or more broadly the Hamiltonian energy eigenvalue problem, is a challenging task, classified as a QMA-hard problem~\cite{kitaev2002classical}. Before the advent of noisy quantum computers~\cite{preskill2018quantum}, classical algorithms such as quantum Monte Carlo (QMC)~\cite{Varney2009Quantum,qin2016benchmark} and density matrix renormalization group (DMRG)~\cite{jiang2023density,white1992density} have had significant success in studying the Fermi-Hubbard model. However, these methods face limitations, including the notorious sign problem in QMC away from half-filling~\cite{Troyer2005computational} and the difficulties in applying DMRG to higher-dimensional systems or systems with periodic boundary conditions~\cite{LeBlanc2015Solutions}.

Recently, algorithms that could run efficiently on quantum computers have also been proposed. These include adiabatic state preparation~\cite{born1928beweis,jansen2007bounds,farhi2000quantum}, shortcuts to adiabaticity~\cite{guery2019shortcuts,hegade2021shortcuts,demirplak2003adiabatic,demirplak2005assisted,berry2009transitionless}, and quantum phase estimation~\cite{Algorithms1994shor,kitaev1995quantum}.  Various variational algorithms have also been proposed, such as the variational quantum eigensolver (VQE)~\cite{mcclean2016theory,Bharti2022Noisy,peruzzo2014variational,cerezo2021variational}, ADAPT-VQE~\cite{grimsley2019adaptive,Gyawali2022adaptive}, global optimization of both parameter values and operator order~\cite{burton2023exact}, feedback-based quantum algorithms~\cite{larsen2024feedback,Magann2022feedback}, variational counterdiabatic techniques~\cite{xie2022variational,claeys2019floquet,tang2024exploring}, artificially engineered cooling systems via an ancilla refrigerator~\cite{marti2024efficient}, and the Hamiltonian variational ansatz (HVA)~\cite{reiner2019finding,wecker2015progress,Cade2020Strategy,cai2020resource}.
We call particular attention to Refs.~\cite{alvertis2025classical,Cade2020Strategy}, which highlight the number-preserving (NP) ansatz~\cite{Cade2020Strategy}, a more accurate generalization of the HVA, and the most accurate wavefunction ansatz prior to this work. However, the performance—specifically, the plateauing error in the ground-state energy per site for larger lattices with strong electronic correlations (\( U/t = 8 \)) remains unsatisfactory, with no clear approach to improve it below 0.01. Moreover, the fidelity can still be substantially below 0.99, such as around 0.7 in the $3\times 3$ case~\cite{alvertis2025classical}.

In this work, we adopt a different approach by employing an engineered cooling algorithm that couples the system to a classical reservoir in the instantaneous-response limit, which we term the \textit{classical reservoir method}. We demonstrate that this method is efficient in terms of quantum circuit complexity and the number of parameters needed, robust against disorder when finding the lowest-energy state in each total spin sector, and --- most importantly --- can consistently achieve $\ge 0.99$ fidelity in all the cases we examined.


\section{Formalism} 
The Hamiltonian for an $N$-site repulsive Fermi-Hubbard model is
\begin{equation}
\hat{H} = -\bar{t} \sum_{\langle i, j \rangle, \sigma} (\hat{c}_{i\sigma}^\dagger \hat{c}_{j\sigma} + \text{c.c.}) + U \sum_i \hat{n}_{i\uparrow} \hat{n}_{i\downarrow},
\label{eq:hubbard_H}
\end{equation}
where $\langle i, j \rangle$ denotes a pair of nearest-neighbor sites. The operators \( \hat{c}_{i\sigma}^\dagger \) and \( \hat{c}_{i\sigma} \) represent the fermionic creation and annihilation operators at site \( i \) for spin \( \sigma \in \{\uparrow, \downarrow\} \). The number operators \( \hat{n}_{i\uparrow} \) and \( \hat{n}_{i\downarrow} \) count the number of spin-up and spin-down electrons at site \( i \). The parameter \( \bar{t} \) represents the hopping strength, which is set to 1, while \( U \) denotes the on-site interaction between electrons with opposite spins. We work in the canonical formalism throughout, with a fixed total particle number and a fixed total \( z \)-component of spin.

\subsection{Algorithm}

We prepare the ground state by cooling the system via the classical reservoir method. There are generically two types of reservoirs that can be used for algorithmic cooling---a quantum reservoir or a classical reservoir. A quantum reservoir has quantum degrees of freedom,
which greatly expands the Hilbert space of the combined system plus reservoir. If the reservoir is noninteracting then the reservoir degrees of freedom can be exactly traced out. If we neglect memory effects, then the system's dynamics can be approximated by a Lindblad master equation~\cite{breuer2002theory}, restoring the simulation to the original system Hilbert space, but now describing the evolution of a density matrix rather than a pure state vector. In this case, the resulting state will generically be mixed. 

In contrast, when using a classical reservoir, which does not add any quantum degrees of freedom, a system initially in a pure state remains in a pure state, which is advantageous for ground-state preparation. However, after tracing out the classical reservoir degrees of freedom, the system is driven by a complex two-time field that creates and annihilates particles, enabling the exchange of particles (and energy) with the classical reservoir at different times. Such a system must be described by Lagrangian time evolution. However, in the instantaneous-response limit, where the particle enters and returns from the classical reservoir at the same time, we can work with Hamiltonian time evolution instead, using a time-dependent Hamiltonian with the classical reservoir arising as additional time-dependent fields that evolve the initial state.

\begin{figure}[ht]
    \begin{centering}
        \includegraphics[width=\columnwidth]{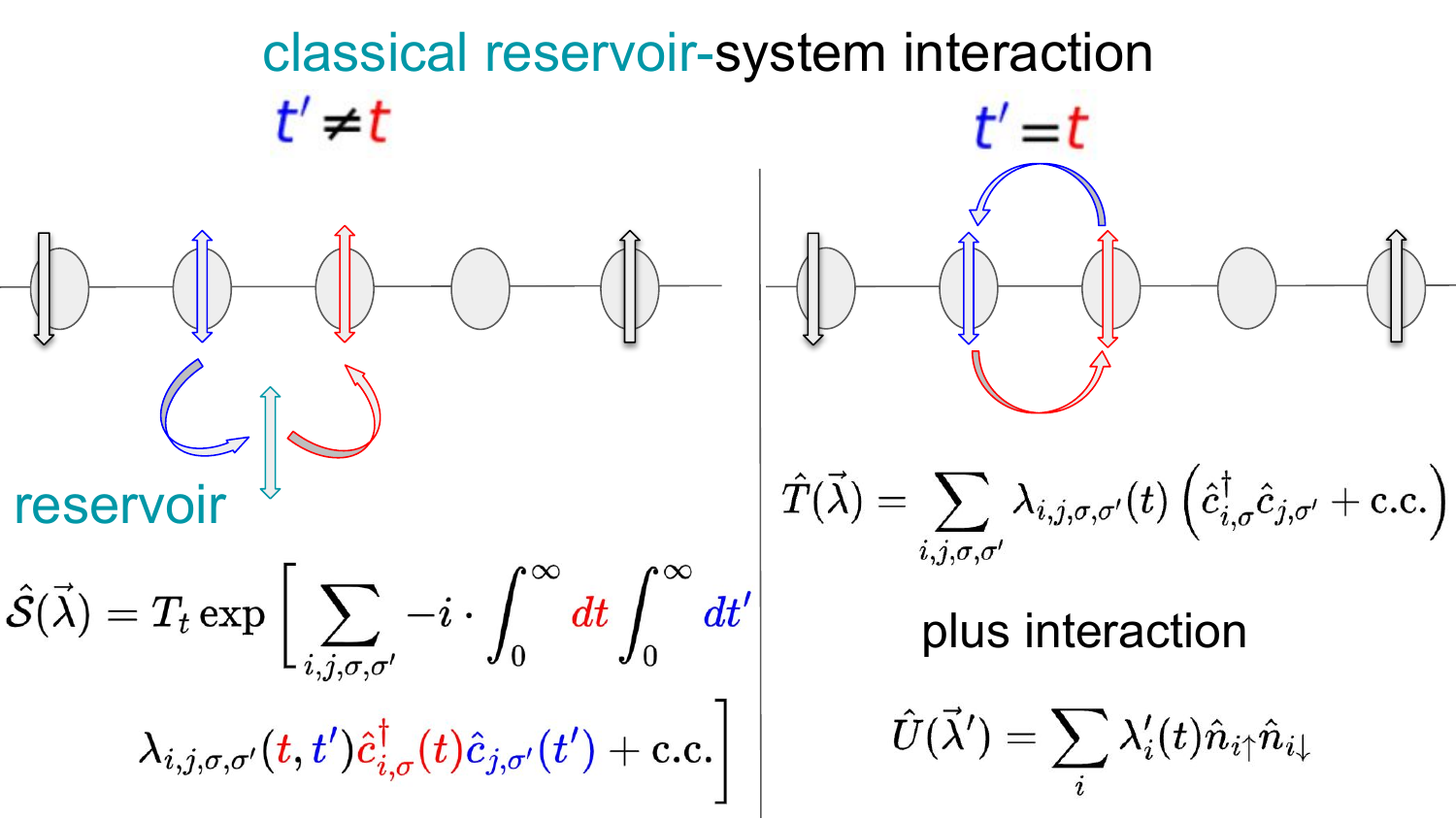}
        \caption{Left, schematic depiction of a generic classical reservoir-system interaction showing the action $\hat{\mathcal{S}}$ describing the retarded (in time) interaction between the reservoir and the system. Right, the instantaneous-response limit used in this work.}
        \label{fig:CR_sceme}
    \end{centering}
\end{figure}

Figure~\ref{fig:CR_sceme} illustrates the transition between Langrangian-based dynamics to Hamiltonian-based dynamics in the instantaneous limit. The Lagrangian framework incorporates the retarded response of the system; i.e. the action \(\hat{\mathcal{S}} \) describes the effects on the system from the reservoir when there is a time lag between particle creation and annihilation at the system-reservoir interface. The reservoir removes particles with spin $\sigma$ from site \( i \) at time \( t \) and reintroduces particles with spin $\sigma^{\prime}$ at site \( j \) at a later time \( t' \), but the classical reservoir does not track the particle dynamics in the reservoir; usually we consider reservoirs that do not flip spins, so $\sigma=\sigma'$. Once \(  \hat{\mathcal{S}} \) is determined, we can calculate the thermal expectation value of physical observables of interest $\hat{O}(t)$ via:
\begin{align}
\langle \hat{O}(t) \rangle &= \frac{1}{Z(\vec{\lambda})}\mathrm{Tr} \left\{ \mathcal{T}_{t} e^{-\beta (\hat{H} - \mu \hat{N}_e)} \hat{\mathcal{S}}(\vec{\lambda}) \hat{O}(t) \right\} \\
Z(\vec{\lambda}) &= \mathrm{Tr} \left\{ \mathcal{T}_{t} e^{-\beta (\hat{H} - \mu \hat{N}_e)} \hat{\mathcal{S}}(\vec{\lambda}) \right\}
\end{align}
where \( \vec{\lambda} \) is the two-time-dependent field describing the coupling between the reservoir and the system, $\mu$ is the chemical potential, \( \hat{N}_e \) is the electron number operator, \( \mathcal{T}_{t} \) is the time-ordering operator, and \( Z(\vec\lambda) \) is the partition function.

\begin{figure*}[htp]
    \begin{centering}
        \includegraphics[width=2\columnwidth]{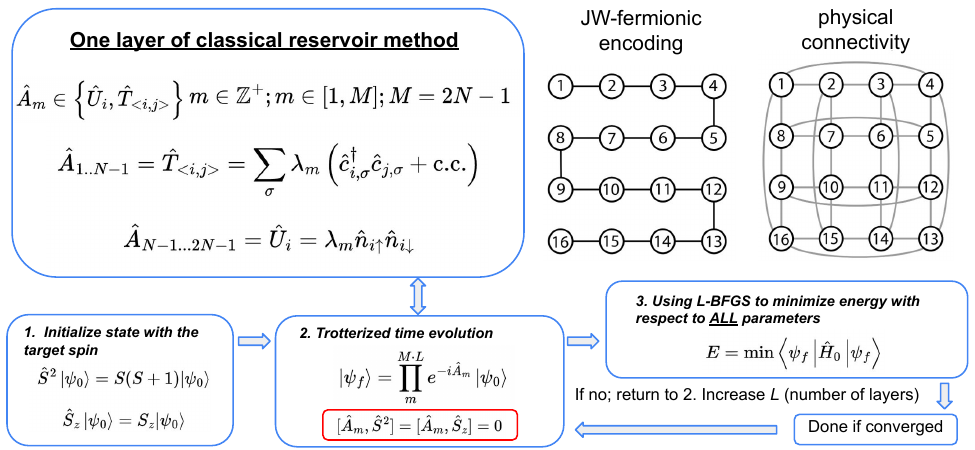}
        \caption{Schematic representation of the classical reservoir method for a \(4 \times 4\) two-dimensional square lattice. Left: operators used to describe the instantaneous-response classical reservoir. Center: snake-like path for the Jordan–Wigner fermionic encoding used in this work, illustrating adjacent sites in the Jordan-Wigner string—i.e., each site has two nearest neighbors in the one-dimensional index scheme, regardless of the actual physical connectivity of the hopping matrix, as shown on the right. Right: connectivity of the hopping terms between nearest neighbors for a small cluster with periodic boundary conditions. Bottom: optimization strategy to prepare the ground state.
        }
        \label{fig:overview_fig}
    \end{centering}
\end{figure*}

There are two major challenges in evaluating \(  \hat{\mathcal{S}} \). First, handling the time ordering is complicated because the  \( \vec{\lambda} \) fields depend on two times, $t$ and $t^{\prime}$. Second, the  grand canonical formalism is required, because particle number is no longer conserved due to the interaction with the reservoir. The dimension increases from the canonical Hilbert-space dimension for $N_e$ electrons on $N$ sites with $S_z=0$, $ \left (\binom{N}{N_e/2}\right )^2$, to the full Hilbert-space dimension of $4^N$ for the system when a generic classical reservoir is incorporated.

The situation greatly simplifies by imposing an instantaneous-response limit on the classical reservoir, where both challenges described above are removed: (i) we can remain in a canonical formalism with fixed particle number and (ii) we can work with a Hamiltonian formalism, where time-ordering is easy to implement because all time-dependent objects depend on only one time. 
The interaction between the system and the reservoir manifests as hopping terms, with the \( \vec{\lambda} \) fields modulating the hopping strengths as shown in Fig.~\ref{fig:CR_sceme}.  Since the exponentials of hopping terms form a closed Lie group, they effectively act as a modification of the single-particle basis. 
To engineer electron correlations, additional operators are required.  We introduce on-site potential terms denoted by \( \hat{U}(\vec{\lambda}^{\prime}) \), where \( \vec{\lambda}^{\prime} \) represents an additional set of time-dependent fields; these fields allow for the creation of entanglement. By alternating between these terms in multiple layers, we construct the ansatz for an effective \quotes{time-evolution operator,}  given by:
\begin{align}
\left|\psi_f\right\rangle = \prod_{l=1}^L e^{-i \Delta t (\hat{T}_l(\vec{\lambda}) + \hat{U}_l(\vec{\lambda}^{\prime}))} \left|\psi_0\right\rangle,
\label{eq:general_CR_state}
\end{align}
where \( |\psi_0\rangle \) is the initial state and \( L \) is the number of layers (analogous to the number of time steps in time evolution). We omit \( \Delta t \) as it can be absorbed into the  \( \vec{\lambda} \) and  \( \vec{\lambda}^{\prime} \) fields because only the product of the two will be optimized.

To make Eq.~(\ref{eq:general_CR_state}) more resource-efficient, three simplifications are applied to derive the final expression for the ansatz. First, the ansatz enforces the conservation of both the total spin and its \( z \)-component, thereby reducing the optimization landscape to a smaller target spin subspace and eliminating concerns about spin contamination. This is achieved by pairing the spin-up and spin-down hopping terms, as previously shown in Ref.~\cite{izmaylov2020order}. Consequently, the initial state must already have the target total spin and \( z \)-component. For a total spin \( S \) state with \( N \) electrons distributed across \( N \) sites (half filling), the configuration of the initial state is as follows: \( \tfrac{N}{2} - S \) empty sites, \( \tfrac{N}{2} - S \) doubly occupied sites, and \( 2S \) spin-up occupied sites. See Appendix A for numerical examples.

Second, we note that the aforementioned closure of the Lie algebra for the singles hopping terms results
\begin{equation}
\left[ \hat{c}_p^{\dagger} \hat{c}_q,\, \hat{c}_a^{\dagger} \hat{c}_b \right] = \hat{c}_p^{\dagger} \delta_{q a} \hat{c}_b - \hat{c}_a^{\dagger} \delta_{b p} \hat{c}_q,
\label{eq:single_commutator}
\end{equation}
where \( p, q, a, b \) are site indices. This implies that including the hopping terms only along the Jordan-Wigner string will create indirect hopping terms between non adjacent sites. This closure property allows us to restrict the operator set to only adjacent hoppings along the Jordan-Wigner snake-like path, regardless of the system's dimension or lattice connectivity when we have enough layers in the ansatz. This reduces computational costs by avoiding the need for any Jordan–Wigner Pauli-\( Z \) string operators in the state preparation~\cite{jordan1993paulische}. Consequently, the hopping part in the ansatz is further simplified to
\begin{equation}
\hat{T}(\vec{\lambda}) = \sum_{\langle i,j \rangle, \sigma} \lambda_{\langle i,j \rangle}(t) \left( \hat{c}_{i,\sigma}^{\dagger} \hat{c}_{j,\sigma} + \textrm{c.c.} \right),
\label{eq:spin_hop}
\end{equation}
where \( \langle i,j \rangle \) denotes adjacent sites along the 1D Jordan–Wigner snake-like path. Conceptually, this is akin to the compression algorithms for free fermionic systems discussed in Refs~\cite{kokcu2022algebraic,kokcu2023algebraic}.

Finally, we separate the hopping and potential terms for efficient implementation on a quantum computer. When separating the hopping terms, one can combine terms that commute into one exponent. In the case of an even number of sites, these terms are separated into two groups. The first group includes hopping terms with hopping indices such as \( \langle 1,2 \rangle \), \( \langle 3,4 \rangle \) ...\( \langle N-1,N \rangle \), and so on, while the second group consists of terms such as \( \langle 2,3 \rangle \), \( \langle 4,5 \rangle \)... \( \langle N-2,N-1 \rangle \), and so forth. Denoting the first hopping group as \( \hat{T}(\vec{\lambda}) \) and the second hopping group as \( \hat{T}^{\prime}(\vec{\lambda}) \), we can rewrite the classical reservoir method ansatz as
\begin{align}
\left|\psi_f\right\rangle &= \prod_{l=1}^{L} e^{i \hat{U}_l(\vec{\lambda}^{\prime})} e^{-i \hat{T}_l(\vec{\lambda})} e^{i \hat{T}^{\prime}_l(\vec{\lambda})}  \left|\psi_0\right\rangle ,\\
&= \prod_{m=1}^{M \cdot L} e^{-i \hat{A}_m} \left|\psi_0\right\rangle,
\label{eq:final_state}
\end{align}
with \( m \) being the parameter index of a set $A$ that combines all the terms in the ansatz. The whole classical reservoir method is summarized in Fig.~\ref{fig:overview_fig} as a flow chart, outlining the process from state initialization and the ansatz expression in Eq.~\ref{eq:final_state}, as well as the fermionic encoding used to translate the problem onto quantum circuits via the Jordan-Wigner transformation. In detail, one ansatz layer contains a total of \( M = 2N - 1 \) parameters, and each ansatz layer consists of three quantum circuit elements each implemented in parallel (on an all-to-all connected machine). We then perform optimizations on the \( \lambda \) and \( \lambda^{\prime} \) fields using noiseless energy measurements (optimization details are provided in Appendix B).

\subsection{Quantum circuit}

To translate the algorithm into a quantum circuit—i.e., to map fermionic operators to Pauli operators—we apply the Jordan–Wigner transformation. Below, we write the on-site potential term and the spin-adapted nearest-neighbor hopping term of the Hubbard Hamiltonian after this mapping, as these are the only operators used in the ansatz.

\begin{equation}
n_{i\uparrow} n_{i\downarrow} \mapsto \frac{1}{4} (I - Z_{i})(I - Z_{i+N}),
\end{equation}
and
\begin{align}
&c_{i\uparrow}^\dagger c_{i+1\uparrow} + c_{i+1\uparrow}^\dagger c_{i\uparrow} + c_{i\downarrow}^\dagger c_{i+1\downarrow} + c_{i+1\downarrow}^\dagger c_{i\downarrow} \nonumber \\
&\mapsto \frac{1}{2} (X_i X_{i+1} + Y_i Y_{i+1} + X_{i+N} X_{i+1+N} + Y_{i+N} Y_{i+1+N}).
\label{eq:hopping}
\end{align}
where \(i\) is the site index. Here, the first \(N\) qubits represent the spin-up orbitals, and the second \(N\) qubits represent the spin-down orbitals. Note that in Eq.~\ref{eq:hopping}, by including only nearest-neighbor terms in the Jordan–Wigner ordering for the ansatz, no \(Z\) string appears in the  quantum state preparation circuit. However, in the measurement circuits, simulating physical Hubbard hopping terms that are not adjacent in the Jordan–Wigner mapping still requires the inclusion of \(Z\) strings to preserve the correct fermionic anticommutation relations.

\begin{figure}
  \centering
  \includegraphics[width=\columnwidth]{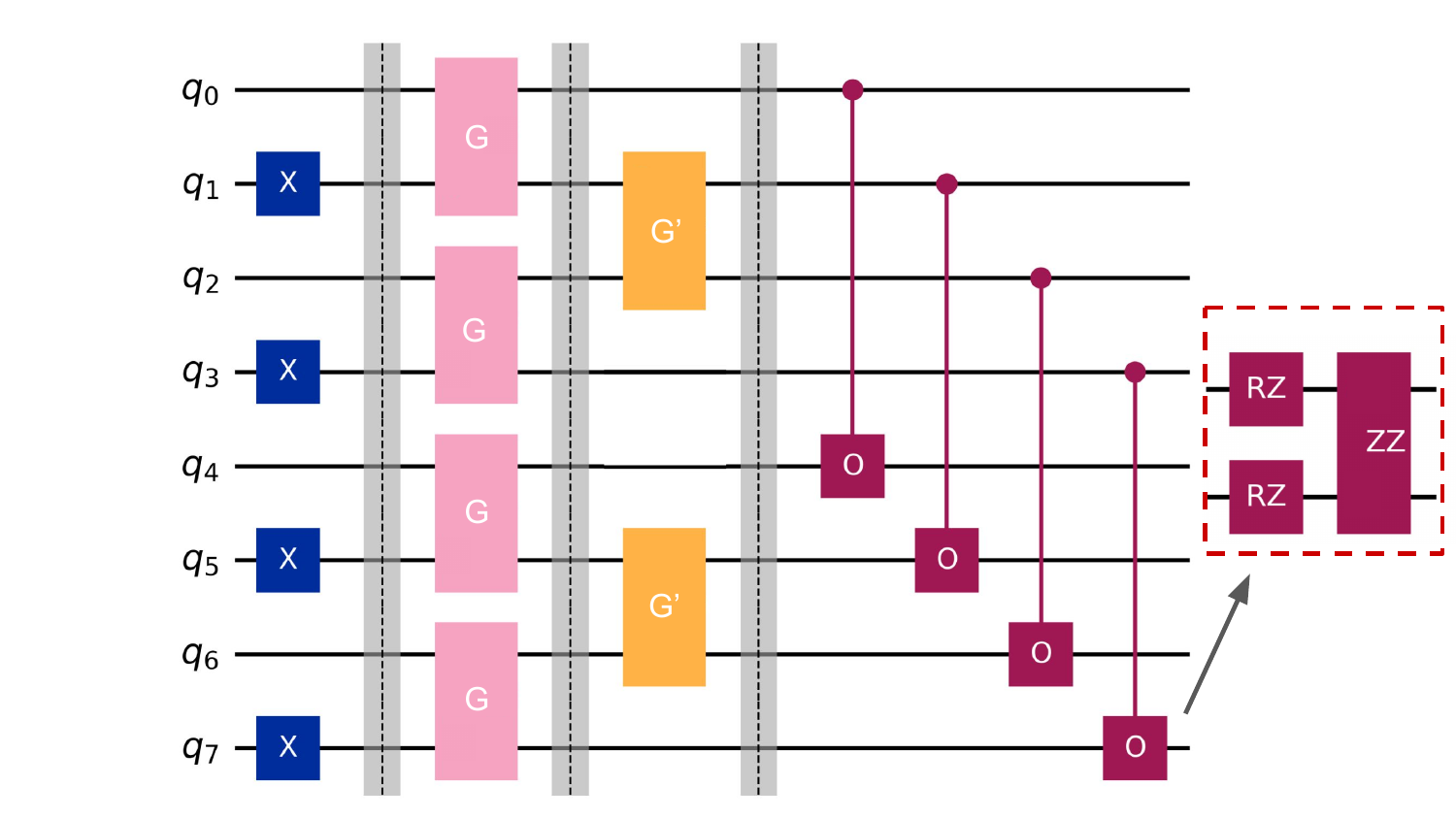}
  \caption{A four-site Hubbard model example with a single layer of the classical reservoir ansatz, including initial state preparation. Qubits \(q_0\) to \(q_3\) denote the spin-up orbitals, while the remaining qubits represent the spin-down orbitals. On the right, we show one possible decomposition of the on-site potential terms into basic quantum gates: two single-qubit rotation \(Z\) gate and a two-qubit \(RZZ\) gate. }
  \label{fig:hubbard circuit}
\end{figure}

In Fig.~\ref{fig:hubbard circuit}, we present a quantum circuit diagram where each color denotes a different layer used in the ansatz. The first single layer of blue \(X\) gates is used for initial state preparation—for example, to create a double occupancy state when preparing the ground state in the \(S=0\) sector. This is followed by two separate layers for nearest-neighbor hopping terms—represented by with pink and orange, corresponding to \( \hat{T}(\vec{\lambda}) \) and \( \hat{T}'(\vec{\lambda}) \), respectively. These layers are denoted by the symbol \(G\) and  \(G'\) , as Givens rotation gates either are the typical native gates on some quantum hardware platforms~\cite{qiskit2024,Cirq_Developers_2025}, or readily map onto them. A final layer for the on-site potential terms, corresponding to the \(ZZ\) interaction is labeled in red. Hence, it is clear that each ansatz layer consists of three quantum circuit layers on fully connected hardware. 
On machines with linear connectivity, the hopping terms remain 2-local interactions. The long-range entangling gates required for the on-site potential terms can be implemented efficiently using dynamic circuits, as proposed in Ref.~\cite{baumer2024efficient}. Alternatively, on the Google Willow chips~\cite{acharya2024quantum}, which feature four-way connectivity in the bulk, one can arrange the spin-up orbitals along one linear array and the spin-down orbitals along another. Since these two linear arrays are positioned adjacent to each other, the on-site potential terms can be implemented as nearest-neighbor gates along the vertical direction in the qubit layout.

\section{Numerical results}

To quantify the accuracy of this work, we define the relative error in energy as \( (E_{gs} - E) / E_{gs} \) and also use the infidelity as a complementary metric. In this work, we examine only two-dimensional systems due to memory constraints in the classical simulation. However, we study varying lattice structures to best illustrate that this method shows promise for application to arbitrary Hubbard models.

\subsection{Two-dimensional periodic cluster }
\begin{figure}[htpb]
    \begin{centering}
        \includegraphics[width=\columnwidth]{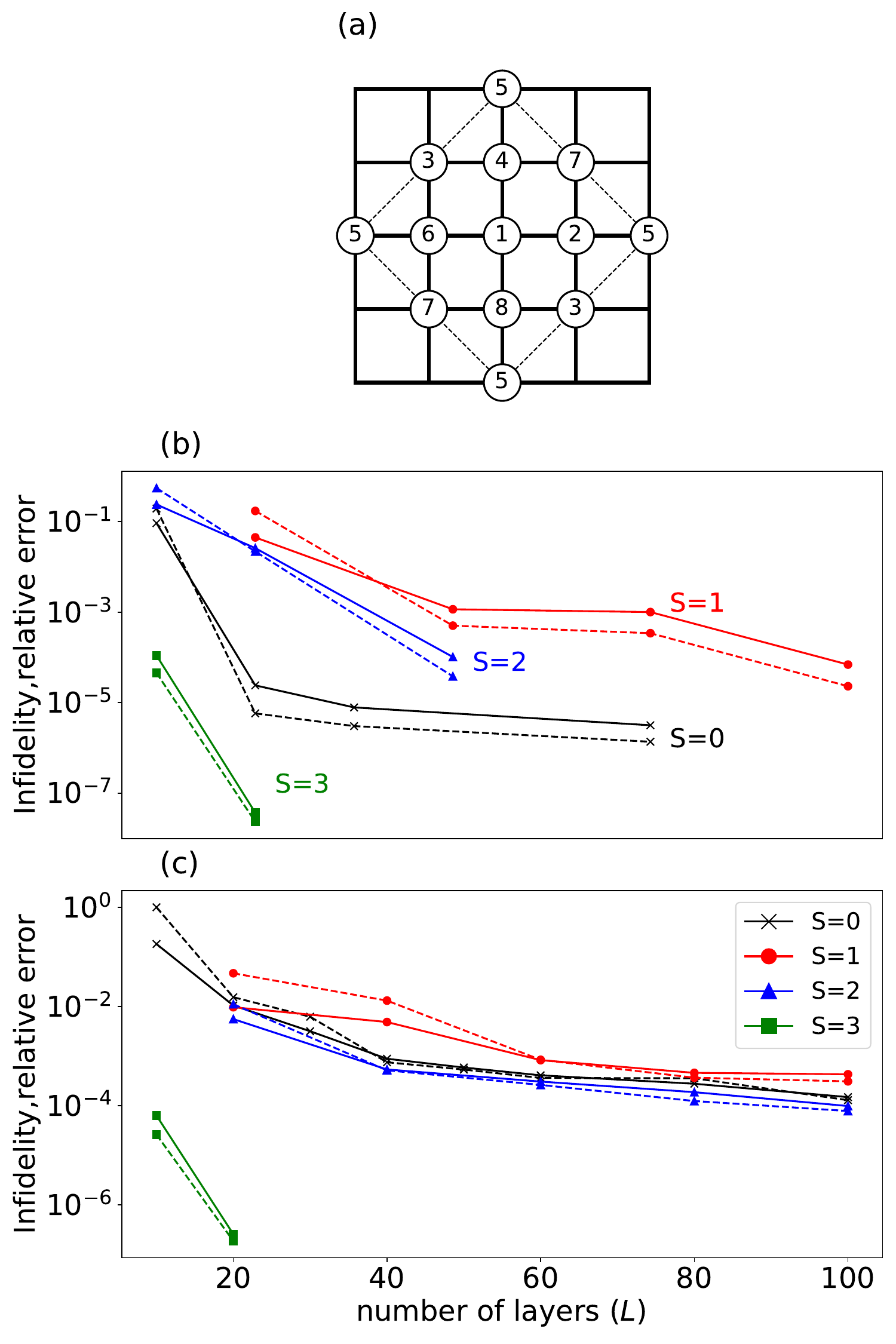}
        \caption{Numerical results for \( U/t = 8 \) in the two-dimensional periodic cluster lattice, with the x-axis representing the number of layers \( L \), where each layer consists of 15 parameters. The top panel (a) illustrates the lattice connectivity. Panel (b) presents the relative error (solid lines) and infidelity (dashed lines) for the pure system, while panel (c) shows the corresponding results for the disordered system. The legend indicates different values of the total spin \( S \), with \( S = 0 \) representing the ground state across all spin configurations in both panels (b) and (c).}
        \label{fig:8_sites_square_lattice_plot}
    \end{centering}
\end{figure}

We perform calculations on a two-dimensional \( \sqrt{8} \times \sqrt{8} \) periodic cluster~\cite{betts1996improved}, where each even-numbered site is connected to four neighboring odd-numbered sites, and vice versa, as illustrated in Fig.~\ref{fig:8_sites_square_lattice_plot} (top panel). When studying two-dimensional systems, this model is interesting because it highlights the potential advantage of quantum computing, as it exhibits faster entanglement growth compared to the two-leg ladder system, which can typically be solved efficiently using DMRG classically~\cite{noack1996ground}. 
 
We can efficiently determine the ground state for different total spin values by starting with an initial state that has the target spin value. The performance for varying numbers of layers at each spin value is detailed in Fig.~\ref{fig:8_sites_square_lattice_plot}, panel (b). Naturally, we find that the difficulty of finding the ground state for different total spin values is directly related to the energy gap in each total spin sector, which is shown in Table~\ref{table:energy_gap}, with a smaller gap making it harder to find the ground state. For the pure Hubbard Hamiltonian at \( U/t = 8 \), which is the case studied in this work, we found the largest energy gap occurs at $S =3$, followed by $S = 0$, then $S =2$, and finally the smallest gap occurs for $S = 1$.  We similarly see that the $S=3$ ground state is the easiest to prepare and the ease of  preparing the ground state increases as the size of the energy gap in each spin sector, as expected, and shown in Fig.~\ref{fig:8_sites_square_lattice_plot} (b). 
\begin{table}
    \centering
    \renewcommand{\arraystretch}{1.2} 
    \begin{tabular}{|c|c|c|c|c|}
        \hline
         & \( S = 0 \) & \( S = 1 \) & \( S = 2 \) & \( S = 3 \) \\ \hline
        pure   & 1.12945  & 0.62746  & 0.78693  & 1.63385  \\ \hline
        disordered & 0.56782  & 0.16508  & 0.32210  & 2.12290  \\ \hline
    \end{tabular}
    \caption{Energy gap values (in units of \( \bar{t} \)) for the pure and and one explicit representation of a disordered Hubbard Hamiltonians for different total spin \( S \) values at $U/t =8$.
}
    \label{table:energy_gap}
\end{table}

We also examine a disordered Hubbard Hamiltonian. The hopping term in Eq.~\ref{eq:hubbard_H} is now given by \( \bar{t}_{ij} = \bar{t} + \delta_{ij} \), where \( \delta_{ij} = 0.2 \cdot \mathcal{N}(0, 1) \), and \( \mathcal{N}(0, 1) \) represents a random number drawn from a normalized Gaussian distribution with mean zero and variance one. The on-site potential term \( U_i \) is a random value drawn from the range \([0, 16]\). In Fig.~\ref{fig:8_sites_square_lattice_plot}, panel (c), it shows that the disorder does not significantly affect performance, suggesting robustness against randomness and against smaller energy gaps in the different spin sectors.

\subsection{Two leg ladder system }

\begin{table*}
  \centering
  \begin{tabular}{@{}lccccc@{}}
    \toprule
    \textbf{Ansatz} 
      & \textbf{Fermionic swap network} 
      & \textbf{Initial state preparation layers} 
      & \textbf{\# Params (2$\times$4)} 
      & \textbf{\# Params (2$\times$5)} \\
    \midrule
    NP (Ref.~\cite{Cade2020Strategy}) 
      & Yes & $N-1$ & 392 & 432 \\
    This work 
      & No  & 1      & 180 & 285 \\
    \bottomrule
  \end{tabular}
  \caption{Resources required to achieve at least 0.99 fidelity in a \(2 \times 4\) and \(2 \times 5\) Hubbard ladder with open boundary conditions and \(U/t = 2\). In the NP ansatz~\cite{Cade2020Strategy}, the initial state preparation requires \(N-1\) quantum circuit layers, which is consistent with the result in Ref.~\cite{kivlichan2018quantum} stating that at most \(N\) layers are needed. In this work, the initial state preparation consists of a single layer of \( X \) gates applied to a subset of qubits, used to encode the desired spin configuration. }
  \label{table: compare}
\end{table*}

We also perform calculations on a two-leg ladder system to benchmark with the number preserving (NP) ansatz~\cite{Cade2020Strategy}. When using the NP ansatz~\cite{Cade2020Strategy}, it is unclear how to handle periodic boundary conditions efficiently~\cite{Cade2020Strategy,kivlichan2018quantum}. Therefore, the comparison between this work and the NP ansatz~\cite{Cade2020Strategy} is conducted using open boundary conditions, with the results from this work summarized in Table~\ref{table: compare}.

\begin{figure}[htbp]
    \begin{centering}
        \includegraphics[width=\columnwidth]{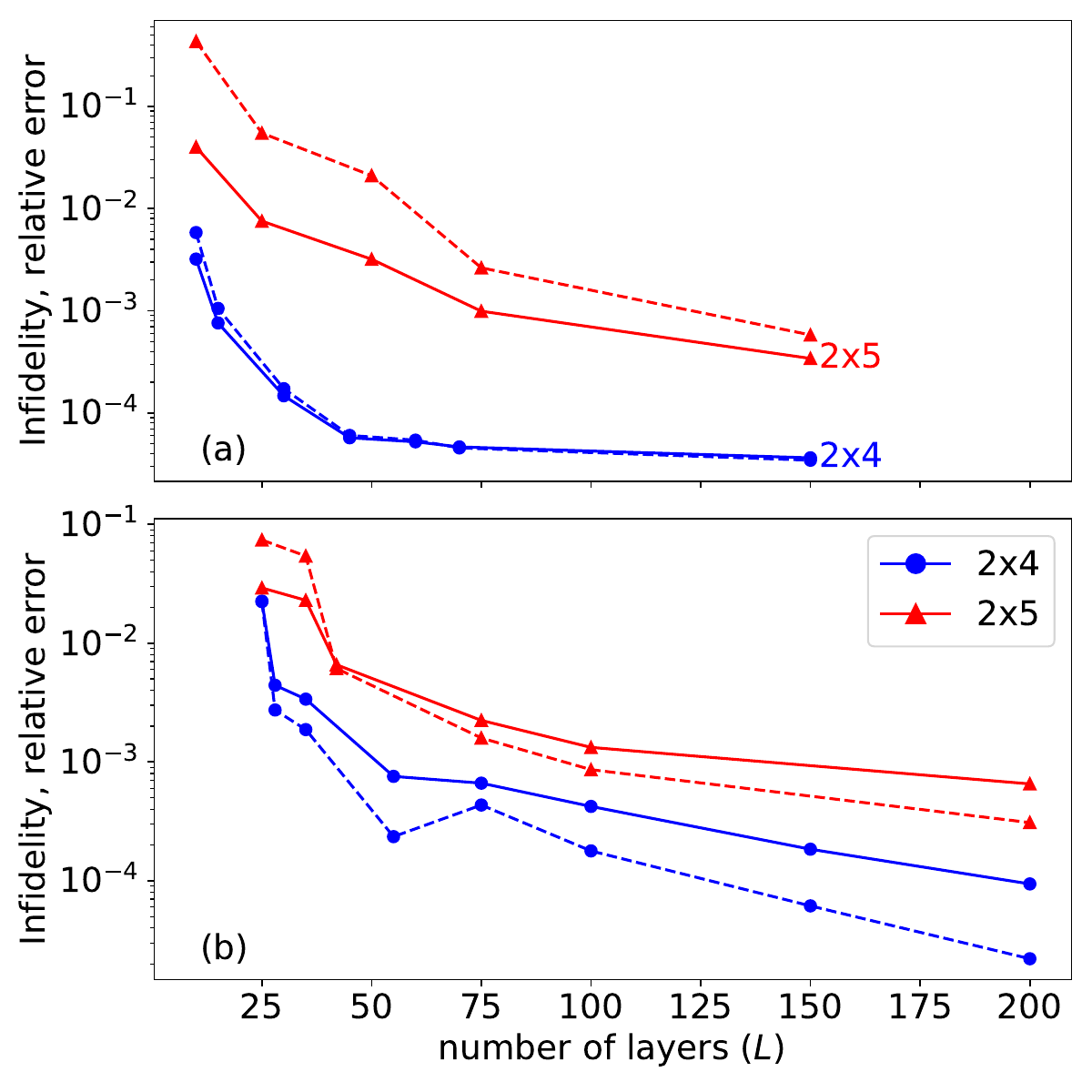}
        \caption{Relative errors (solid lines) and infidelities (dashed lines) for the two-leg ladder system with 8 sites and 10 sites. Panel (a) corresponds to \( U/t = 2 \), while panel (b) corresponds to \( U/t = 8 \).}
        \label{fig:two_leg}
    \end{centering}
\end{figure}

Systems with periodic boundary conditions typically have reduced finite-size effects, because they have no surface effects~\cite{gros1996control}. In Fig.~\ref{fig:two_leg}, we present numerical results for \(2 \times 4\) and \(2 \times 5\) ladders in both the weakly and strongly correlated regimes, using periodic boundary conditions. Notably, by simply adding more parameters and layers, the ansatz can achieve accurate ground-state preparation, with fidelities reaching at least 0.999.

From the ladder system data, we see that this method offers two key improvements. First, this work successfully prepares the ground state in the strongly correlated regime with a high fidelity.
Second, this work demonstrates a potential pathway for efficiently simulating multi-dimensional systems by employing the one-dimensional indexing scheme of the Jordan-Wigner string. An intriguing observation arises when comparing the \(2 \times 4\) ladder system to a two-dimensional periodic cluster: the latter, with increased site connectivity, surprisingly yields higher final fidelities. This suggests that the classical reservoir approach may be particularly well-suited for simulating realistic lattices with higher connectivity.

\section{Conclusion}

We show that the simplification to an instantaneous interaction with a classical reservoir is an efficient method for ground-state preparation, offering a  feasible approach for implementation in near-term quantum computers.

When comparing this work to previous approaches, we find that the classical reservoir method produces an ansatz expression analogous to the NP ansatz or the HVA ansatz~\cite{Cade2020Strategy,wecker2015progress}. However, the motivations of the two approaches are different---our method arises from examining cooling dynamics while the HVA and NP ansatz are motivated by the adiabatic theorem~\cite{Cade2020Strategy,wecker2015progress}. 
The classical reservoir method further enhances the efficiency of the algorithm by exploiting total spin symmetry and the closure property of the Lie algebra in the Hubbard hopping terms. Depending on hardware constraints, our algorithm can be implemented in a similar fashion to NP ansatz ~\cite{Cade2020Strategy,alvertis2025classical}.

The classical reservoir method shows promise in preparing more accurate ground states in the strongly correlated regime, while simultaneously reducing implementation costs in the following ways:
\begin{enumerate}
\item We only require the initial state to have the target total spin and do not need it to be the \( U{=}0 \) non-interacting ground state; it is most easily prepared by creating double occupancies in a single product state. The noininteracting state is more complicated to make in the position basis~\cite{Cade2020Strategy,Jiang2018quantum}, especially when it is degenerate~\cite{wecker2015progress}. 
\item Constraining the system to a definite total spin as well as its $z$-component allows us to find the ground state for each value of total spin easily and reduces the numerical cost by working within a smaller optimization landscape.
\item The closure property of the Lie algebra allows us to use a one-dimensional indexing scheme, eliminating the need for fermionic swap (fSWAP) network~\cite{Verstraete2009quantum,kivlichan2018quantum} to handle any hopping terms that require Jordan-Wigner strings in the ansatz for the state preparation.
\end{enumerate}

Using the classical reservoir method, we find high-accuracy results ---infidelities reach the range of \( 10^{-3} - 10^{-6} \) for 8- and 10-site two-dimensional systems, even in the strong correlation region (\( U/t{=}8 \)). We conjecture that this robustness arises from the fact that we do not start with the $U=0$ ground state as many other algorithms do, and thereby all $U$ values are essentially equal in terms of preparation difficulty.

\section{Acknowledgments }
This work was supported by the Department of Energy, Office of Basic Energy Sciences, Division of Materials Sciences and Engineering under grant no. DE-SC0023231. J.K.F. was also supported by the McDevitt bequest at Georgetown.
\section{Data Availability}

The data that support the findings of this article as well as the python code that run the calculations are openly available at~\cite{data}. 

\bibliographystyle{apsrev4-2}
\bibliography{bib.bib}

\begin{thebibliography}{56}%
\makeatletter
\providecommand \@ifxundefined [1]{%
 \@ifx{#1\undefined}
}%
\providecommand \@ifnum [1]{%
 \ifnum #1\expandafter \@firstoftwo
 \else \expandafter \@secondoftwo
 \fi
}%
\providecommand \@ifx [1]{%
 \ifx #1\expandafter \@firstoftwo
 \else \expandafter \@secondoftwo
 \fi
}%
\providecommand \natexlab [1]{#1}%
\providecommand \enquote  [1]{``#1''}%
\providecommand \bibnamefont  [1]{#1}%
\providecommand \bibfnamefont [1]{#1}%
\providecommand \citenamefont [1]{#1}%
\providecommand \href@noop [0]{\@secondoftwo}%
\providecommand \href [0]{\begingroup \@sanitize@url \@href}%
\providecommand \@href[1]{\@@startlink{#1}\@@href}%
\providecommand \@@href[1]{\endgroup#1\@@endlink}%
\providecommand \@sanitize@url [0]{\catcode `\\12\catcode `\$12\catcode
  `\&12\catcode `\#12\catcode `\^12\catcode `\_12\catcode `\%12\relax}%
\providecommand \@@startlink[1]{}%
\providecommand \@@endlink[0]{}%
\providecommand \url  [0]{\begingroup\@sanitize@url \@url }%
\providecommand \@url [1]{\endgroup\@href {#1}{\urlprefix }}%
\providecommand \urlprefix  [0]{URL }%
\providecommand \Eprint [0]{\href }%
\providecommand \doibase [0]{https://doi.org/}%
\providecommand \selectlanguage [0]{\@gobble}%
\providecommand \bibinfo  [0]{\@secondoftwo}%
\providecommand \bibfield  [0]{\@secondoftwo}%
\providecommand \translation [1]{[#1]}%
\providecommand \BibitemOpen [0]{}%
\providecommand \bibitemStop [0]{}%
\providecommand \bibitemNoStop [0]{.\EOS\space}%
\providecommand \EOS [0]{\spacefactor3000\relax}%
\providecommand \BibitemShut  [1]{\csname bibitem#1\endcsname}%
\let\auto@bib@innerbib\@empty
\bibitem [{\citenamefont {Kitaev}\ \emph {et~al.}(2002)\citenamefont {Kitaev},
  \citenamefont {Shen},\ and\ \citenamefont {Vyalyi}}]{kitaev2002classical}%
  \BibitemOpen
  \bibfield  {author} {\bibinfo {author} {\bibfnamefont {A.~Y.}\ \bibnamefont
  {Kitaev}}, \bibinfo {author} {\bibfnamefont {A.}~\bibnamefont {Shen}},\ and\
  \bibinfo {author} {\bibfnamefont {M.~N.}\ \bibnamefont {Vyalyi}},\
  }\href@noop {} {\emph {\bibinfo {title} {Classical and quantum
  computation}}},\ \bibinfo {number} {47}\ (\bibinfo  {publisher} {American
  Mathematical Soc.},\ \bibinfo {year} {2002})\BibitemShut {NoStop}%
\bibitem [{\citenamefont {Preskill}(2018)}]{preskill2018quantum}%
  \BibitemOpen
  \bibfield  {author} {\bibinfo {author} {\bibfnamefont {J.}~\bibnamefont
  {Preskill}},\ }\href
  {https://doi.org/https://doi.org/10.22331/q-2018-08-06-79} {\bibfield
  {journal} {\bibinfo  {journal} {Quantum}\ }\textbf {\bibinfo {volume} {2}},\
  \bibinfo {pages} {79} (\bibinfo {year} {2018})}\BibitemShut {NoStop}%
\bibitem [{\citenamefont {Varney}\ \emph {et~al.}(2009)\citenamefont {Varney},
  \citenamefont {Lee}, \citenamefont {Bai}, \citenamefont {Chiesa},
  \citenamefont {Jarrell},\ and\ \citenamefont
  {Scalettar}}]{Varney2009Quantum}%
  \BibitemOpen
  \bibfield  {author} {\bibinfo {author} {\bibfnamefont {C.~N.}\ \bibnamefont
  {Varney}}, \bibinfo {author} {\bibfnamefont {C.-R.}\ \bibnamefont {Lee}},
  \bibinfo {author} {\bibfnamefont {Z.~J.}\ \bibnamefont {Bai}}, \bibinfo
  {author} {\bibfnamefont {S.}~\bibnamefont {Chiesa}}, \bibinfo {author}
  {\bibfnamefont {M.}~\bibnamefont {Jarrell}},\ and\ \bibinfo {author}
  {\bibfnamefont {R.~T.}\ \bibnamefont {Scalettar}},\ }\href
  {https://doi.org/10.1103/PhysRevB.80.075116} {\bibfield  {journal} {\bibinfo
  {journal} {Phys. Rev. B}\ }\textbf {\bibinfo {volume} {80}},\ \bibinfo
  {pages} {075116} (\bibinfo {year} {2009})}\BibitemShut {NoStop}%
\bibitem [{\citenamefont {Qin}\ \emph {et~al.}(2016)\citenamefont {Qin},
  \citenamefont {Shi},\ and\ \citenamefont {Zhang}}]{qin2016benchmark}%
  \BibitemOpen
  \bibfield  {author} {\bibinfo {author} {\bibfnamefont {M.}~\bibnamefont
  {Qin}}, \bibinfo {author} {\bibfnamefont {H.}~\bibnamefont {Shi}},\ and\
  \bibinfo {author} {\bibfnamefont {S.}~\bibnamefont {Zhang}},\ }\href
  {https://doi.org/10.1103/PhysRevB.94.085103} {\bibfield  {journal} {\bibinfo
  {journal} {Phys. Rev. B}\ }\textbf {\bibinfo {volume} {94}},\ \bibinfo
  {pages} {085103} (\bibinfo {year} {2016})}\BibitemShut {NoStop}%
\bibitem [{\citenamefont {Jiang}\ \emph {et~al.}(2023)\citenamefont {Jiang},
  \citenamefont {Scalapino},\ and\ \citenamefont {White}}]{jiang2023density}%
  \BibitemOpen
  \bibfield  {author} {\bibinfo {author} {\bibfnamefont {S.}~\bibnamefont
  {Jiang}}, \bibinfo {author} {\bibfnamefont {D.~J.}\ \bibnamefont
  {Scalapino}},\ and\ \bibinfo {author} {\bibfnamefont {S.~R.}\ \bibnamefont
  {White}},\ }\href {https://doi.org/10.1103/PhysRevB.108.L161111} {\bibfield
  {journal} {\bibinfo  {journal} {Phys. Rev. B}\ }\textbf {\bibinfo {volume}
  {108}},\ \bibinfo {pages} {L161111} (\bibinfo {year} {2023})}\BibitemShut
  {NoStop}%
\bibitem [{\citenamefont {White}(1992)}]{white1992density}%
  \BibitemOpen
  \bibfield  {author} {\bibinfo {author} {\bibfnamefont {S.~R.}\ \bibnamefont
  {White}},\ }\href {https://doi.org/10.1103/PhysRevLett.69.2863} {\bibfield
  {journal} {\bibinfo  {journal} {Phys. Rev. Lett.}\ }\textbf {\bibinfo
  {volume} {69}},\ \bibinfo {pages} {2863} (\bibinfo {year}
  {1992})}\BibitemShut {NoStop}%
\bibitem [{\citenamefont {Troyer}\ and\ \citenamefont
  {Wiese}(2005)}]{Troyer2005computational}%
  \BibitemOpen
  \bibfield  {author} {\bibinfo {author} {\bibfnamefont {M.}~\bibnamefont
  {Troyer}}\ and\ \bibinfo {author} {\bibfnamefont {U.-J.}\ \bibnamefont
  {Wiese}},\ }\href {https://doi.org/10.1103/PhysRevLett.94.170201} {\bibfield
  {journal} {\bibinfo  {journal} {Phys. Rev. Lett.}\ }\textbf {\bibinfo
  {volume} {94}},\ \bibinfo {pages} {170201} (\bibinfo {year}
  {2005})}\BibitemShut {NoStop}%
\bibitem [{\citenamefont {LeBlanc}\ \emph {et~al.}(2015)\citenamefont
  {LeBlanc}, \citenamefont {Antipov}, \citenamefont {Becca}, \citenamefont
  {Bulik}, \citenamefont {Chan}, \citenamefont {Chung}, \citenamefont {Deng},
  \citenamefont {Ferrero}, \citenamefont {Henderson}, \citenamefont
  {Jim\'enez-Hoyos}, \citenamefont {Kozik}, \citenamefont {Liu}, \citenamefont
  {Millis}, \citenamefont {Prokof'ev}, \citenamefont {Qin}, \citenamefont
  {Scuseria}, \citenamefont {Shi}, \citenamefont {Svistunov}, \citenamefont
  {Tocchio}, \citenamefont {Tupitsyn}, \citenamefont {White}, \citenamefont
  {Zhang}, \citenamefont {Zheng}, \citenamefont {Zhu},\ and\ \citenamefont
  {Gull}}]{LeBlanc2015Solutions}%
  \BibitemOpen
  \bibfield  {author} {\bibinfo {author} {\bibfnamefont {J.~P.~F.}\
  \bibnamefont {LeBlanc}}, \bibinfo {author} {\bibfnamefont {A.~E.}\
  \bibnamefont {Antipov}}, \bibinfo {author} {\bibfnamefont {F.}~\bibnamefont
  {Becca}}, \bibinfo {author} {\bibfnamefont {I.~W.}\ \bibnamefont {Bulik}},
  \bibinfo {author} {\bibfnamefont {G.~K.-L.}\ \bibnamefont {Chan}}, \bibinfo
  {author} {\bibfnamefont {C.-M.}\ \bibnamefont {Chung}}, \bibinfo {author}
  {\bibfnamefont {Y.}~\bibnamefont {Deng}}, \bibinfo {author} {\bibfnamefont
  {M.}~\bibnamefont {Ferrero}}, \bibinfo {author} {\bibfnamefont {T.~M.}\
  \bibnamefont {Henderson}}, \bibinfo {author} {\bibfnamefont {C.~A.}\
  \bibnamefont {Jim\'enez-Hoyos}}, \bibinfo {author} {\bibfnamefont
  {E.}~\bibnamefont {Kozik}}, \bibinfo {author} {\bibfnamefont {X.-W.}\
  \bibnamefont {Liu}}, \bibinfo {author} {\bibfnamefont {A.~J.}\ \bibnamefont
  {Millis}}, \bibinfo {author} {\bibfnamefont {N.~V.}\ \bibnamefont
  {Prokof'ev}}, \bibinfo {author} {\bibfnamefont {M.}~\bibnamefont {Qin}},
  \bibinfo {author} {\bibfnamefont {G.~E.}\ \bibnamefont {Scuseria}}, \bibinfo
  {author} {\bibfnamefont {H.}~\bibnamefont {Shi}}, \bibinfo {author}
  {\bibfnamefont {B.~V.}\ \bibnamefont {Svistunov}}, \bibinfo {author}
  {\bibfnamefont {L.~F.}\ \bibnamefont {Tocchio}}, \bibinfo {author}
  {\bibfnamefont {I.~S.}\ \bibnamefont {Tupitsyn}}, \bibinfo {author}
  {\bibfnamefont {S.~R.}\ \bibnamefont {White}}, \bibinfo {author}
  {\bibfnamefont {S.}~\bibnamefont {Zhang}}, \bibinfo {author} {\bibfnamefont
  {B.-X.}\ \bibnamefont {Zheng}}, \bibinfo {author} {\bibfnamefont
  {Z.}~\bibnamefont {Zhu}},\ and\ \bibinfo {author} {\bibfnamefont
  {E.}~\bibnamefont {Gull}} (\bibinfo {collaboration} {Simons Collaboration on
  the Many-Electron Problem}),\ }\href
  {https://doi.org/10.1103/PhysRevX.5.041041} {\bibfield  {journal} {\bibinfo
  {journal} {Phys. Rev. X}\ }\textbf {\bibinfo {volume} {5}},\ \bibinfo {pages}
  {041041} (\bibinfo {year} {2015})}\BibitemShut {NoStop}%
\bibitem [{\citenamefont {Born}\ and\ \citenamefont
  {Fock}(1928)}]{born1928beweis}%
  \BibitemOpen
  \bibfield  {author} {\bibinfo {author} {\bibfnamefont {M.}~\bibnamefont
  {Born}}\ and\ \bibinfo {author} {\bibfnamefont {V.}~\bibnamefont {Fock}},\
  }\href {https://doi.org/https://doi.org/10.1007/BF01343193} {\bibfield
  {journal} {\bibinfo  {journal} {Z. Phys.}\ }\textbf {\bibinfo {volume}
  {51}},\ \bibinfo {pages} {165} (\bibinfo {year} {1928})}\BibitemShut
  {NoStop}%
\bibitem [{\citenamefont {Jansen}\ \emph {et~al.}(2007)\citenamefont {Jansen},
  \citenamefont {Ruskai},\ and\ \citenamefont {Seiler}}]{jansen2007bounds}%
  \BibitemOpen
  \bibfield  {author} {\bibinfo {author} {\bibfnamefont {S.}~\bibnamefont
  {Jansen}}, \bibinfo {author} {\bibfnamefont {M.-B.}\ \bibnamefont {Ruskai}},\
  and\ \bibinfo {author} {\bibfnamefont {R.}~\bibnamefont {Seiler}},\ }\href
  {https://doi.org/https://doi.org/10.1063/1.2798382} {\bibfield  {journal}
  {\bibinfo  {journal} {J. Math. Phys.}\ }\textbf {\bibinfo {volume} {48}},\
  \bibinfo {pages} {102111} (\bibinfo {year} {2007})}\BibitemShut {NoStop}%
\bibitem [{\citenamefont {Farhi}\ \emph {et~al.}(2000)\citenamefont {Farhi},
  \citenamefont {Goldstone}, \citenamefont {Gutmann},\ and\ \citenamefont
  {Sipser}}]{farhi2000quantum}%
  \BibitemOpen
  \bibfield  {author} {\bibinfo {author} {\bibfnamefont {E.}~\bibnamefont
  {Farhi}}, \bibinfo {author} {\bibfnamefont {J.}~\bibnamefont {Goldstone}},
  \bibinfo {author} {\bibfnamefont {S.}~\bibnamefont {Gutmann}},\ and\ \bibinfo
  {author} {\bibfnamefont {M.}~\bibnamefont {Sipser}},\ }\bibfield  {journal}
  {\bibinfo  {journal} {arXiv}\ }\href
  {https://doi.org/https://doi.org/10.48550/arXiv.quant-ph/0001106}
  {https://doi.org/10.48550/arXiv.quant-ph/0001106} (\bibinfo {year}
  {2000})\BibitemShut {NoStop}%
\bibitem [{\citenamefont {Gu{\'e}ry-Odelin}\ \emph {et~al.}(2019)\citenamefont
  {Gu{\'e}ry-Odelin}, \citenamefont {Ruschhaupt}, \citenamefont {Kiely},
  \citenamefont {Torrontegui}, \citenamefont {Mart{\'\i}nez-Garaot},\ and\
  \citenamefont {Muga}}]{guery2019shortcuts}%
  \BibitemOpen
  \bibfield  {author} {\bibinfo {author} {\bibfnamefont {D.}~\bibnamefont
  {Gu{\'e}ry-Odelin}}, \bibinfo {author} {\bibfnamefont {A.}~\bibnamefont
  {Ruschhaupt}}, \bibinfo {author} {\bibfnamefont {A.}~\bibnamefont {Kiely}},
  \bibinfo {author} {\bibfnamefont {E.}~\bibnamefont {Torrontegui}}, \bibinfo
  {author} {\bibfnamefont {S.}~\bibnamefont {Mart{\'\i}nez-Garaot}},\ and\
  \bibinfo {author} {\bibfnamefont {J.~G.}\ \bibnamefont {Muga}},\ }\href
  {https://doi.org/10.1103/RevModPhys.91.045001} {\bibfield  {journal}
  {\bibinfo  {journal} {Rev. Mod. Phys.}\ }\textbf {\bibinfo {volume} {91}},\
  \bibinfo {pages} {045001} (\bibinfo {year} {2019})}\BibitemShut {NoStop}%
\bibitem [{\citenamefont {Hegade}\ \emph {et~al.}(2021)\citenamefont {Hegade},
  \citenamefont {Paul}, \citenamefont {Ding}, \citenamefont {Sanz},
  \citenamefont {Albarr{\'a}n-Arriagada}, \citenamefont {Solano},\ and\
  \citenamefont {Chen}}]{hegade2021shortcuts}%
  \BibitemOpen
  \bibfield  {author} {\bibinfo {author} {\bibfnamefont {N.~N.}\ \bibnamefont
  {Hegade}}, \bibinfo {author} {\bibfnamefont {K.}~\bibnamefont {Paul}},
  \bibinfo {author} {\bibfnamefont {Y.}~\bibnamefont {Ding}}, \bibinfo {author}
  {\bibfnamefont {M.}~\bibnamefont {Sanz}}, \bibinfo {author} {\bibfnamefont
  {F.}~\bibnamefont {Albarr{\'a}n-Arriagada}}, \bibinfo {author} {\bibfnamefont
  {E.}~\bibnamefont {Solano}},\ and\ \bibinfo {author} {\bibfnamefont
  {X.}~\bibnamefont {Chen}},\ }\href
  {https://doi.org/10.1103/PhysRevApplied.15.024038} {\bibfield  {journal}
  {\bibinfo  {journal} {Phys. Rev. Appl.}\ }\textbf {\bibinfo {volume} {15}},\
  \bibinfo {pages} {024038} (\bibinfo {year} {2021})}\BibitemShut {NoStop}%
\bibitem [{\citenamefont {Demirplak}\ and\ \citenamefont
  {Rice}(2003)}]{demirplak2003adiabatic}%
  \BibitemOpen
  \bibfield  {author} {\bibinfo {author} {\bibfnamefont {M.}~\bibnamefont
  {Demirplak}}\ and\ \bibinfo {author} {\bibfnamefont {S.~A.}\ \bibnamefont
  {Rice}},\ }\href {https://doi.org/https://doi.org/10.1021/jp030708a}
  {\bibfield  {journal} {\bibinfo  {journal} {J. Phys. Chem. A}\ }\textbf
  {\bibinfo {volume} {107}},\ \bibinfo {pages} {9937} (\bibinfo {year}
  {2003})}\BibitemShut {NoStop}%
\bibitem [{\citenamefont {Demirplak}\ and\ \citenamefont
  {Rice}(2005)}]{demirplak2005assisted}%
  \BibitemOpen
  \bibfield  {author} {\bibinfo {author} {\bibfnamefont {M.}~\bibnamefont
  {Demirplak}}\ and\ \bibinfo {author} {\bibfnamefont {S.~A.}\ \bibnamefont
  {Rice}},\ }\href {https://doi.org/10.1088/1751-8113/42/36/365303} {\bibfield
  {journal} {\bibinfo  {journal} {J. Phys. Chem. B}\ }\textbf {\bibinfo
  {volume} {109}},\ \bibinfo {pages} {6838} (\bibinfo {year}
  {2005})}\BibitemShut {NoStop}%
\bibitem [{\citenamefont {Berry}(2009)}]{berry2009transitionless}%
  \BibitemOpen
  \bibfield  {author} {\bibinfo {author} {\bibfnamefont {M.~V.}\ \bibnamefont
  {Berry}},\ }\href {https://doi.org/10.1088/1751-8113/42/36/365303} {\bibfield
   {journal} {\bibinfo  {journal} {J. Phys. A: Math. Theor.}\ }\textbf
  {\bibinfo {volume} {42}},\ \bibinfo {pages} {365303} (\bibinfo {year}
  {2009})}\BibitemShut {NoStop}%
\bibitem [{\citenamefont {Shor}(1994)}]{Algorithms1994shor}%
  \BibitemOpen
  \bibfield  {author} {\bibinfo {author} {\bibfnamefont {P.}~\bibnamefont
  {Shor}},\ }in\ \href {https://doi.org/10.1109/SFCS.1994.365700} {\emph
  {\bibinfo {booktitle} {Proceedings 35th Annual Symposium on Foundations of
  Computer Science}}}\ (\bibinfo {year} {1994})\ pp.\ \bibinfo {pages}
  {124--134}\BibitemShut {NoStop}%
\bibitem [{\citenamefont {Kitaev}(1995)}]{kitaev1995quantum}%
  \BibitemOpen
  \bibfield  {author} {\bibinfo {author} {\bibfnamefont {A.~Y.}\ \bibnamefont
  {Kitaev}},\ }\bibfield  {journal} {\bibinfo  {journal} {arXiv}\ }\href
  {https://doi.org/https://doi.org/10.48550/arXiv.quant-ph/9511026}
  {https://doi.org/10.48550/arXiv.quant-ph/9511026} (\bibinfo {year}
  {1995})\BibitemShut {NoStop}%
\bibitem [{\citenamefont {McClean}\ \emph {et~al.}(2016)\citenamefont
  {McClean}, \citenamefont {Romero}, \citenamefont {Babbush},\ and\
  \citenamefont {Aspuru-Guzik}}]{mcclean2016theory}%
  \BibitemOpen
  \bibfield  {author} {\bibinfo {author} {\bibfnamefont {J.~R.}\ \bibnamefont
  {McClean}}, \bibinfo {author} {\bibfnamefont {J.}~\bibnamefont {Romero}},
  \bibinfo {author} {\bibfnamefont {R.}~\bibnamefont {Babbush}},\ and\ \bibinfo
  {author} {\bibfnamefont {A.}~\bibnamefont {Aspuru-Guzik}},\ }\href
  {https://doi.org/10.1088/1367-2630/18/2/023023} {\bibfield  {journal}
  {\bibinfo  {journal} {New J. Phys.}\ }\textbf {\bibinfo {volume} {18}},\
  \bibinfo {pages} {023023} (\bibinfo {year} {2016})}\BibitemShut {NoStop}%
\bibitem [{\citenamefont {Bharti}\ \emph {et~al.}(2022)\citenamefont {Bharti},
  \citenamefont {Cervera-Lierta}, \citenamefont {Kyaw}, \citenamefont {Haug},
  \citenamefont {Alperin-Lea}, \citenamefont {Anand}, \citenamefont {Degroote},
  \citenamefont {Heimonen}, \citenamefont {Kottmann}, \citenamefont {Menke},
  \citenamefont {Mok}, \citenamefont {Sim}, \citenamefont {Kwek},\ and\
  \citenamefont {Aspuru-Guzik}}]{Bharti2022Noisy}%
  \BibitemOpen
  \bibfield  {author} {\bibinfo {author} {\bibfnamefont {K.}~\bibnamefont
  {Bharti}}, \bibinfo {author} {\bibfnamefont {A.}~\bibnamefont
  {Cervera-Lierta}}, \bibinfo {author} {\bibfnamefont {T.~H.}\ \bibnamefont
  {Kyaw}}, \bibinfo {author} {\bibfnamefont {T.}~\bibnamefont {Haug}}, \bibinfo
  {author} {\bibfnamefont {S.}~\bibnamefont {Alperin-Lea}}, \bibinfo {author}
  {\bibfnamefont {A.}~\bibnamefont {Anand}}, \bibinfo {author} {\bibfnamefont
  {M.}~\bibnamefont {Degroote}}, \bibinfo {author} {\bibfnamefont
  {H.}~\bibnamefont {Heimonen}}, \bibinfo {author} {\bibfnamefont {J.~S.}\
  \bibnamefont {Kottmann}}, \bibinfo {author} {\bibfnamefont {T.}~\bibnamefont
  {Menke}}, \bibinfo {author} {\bibfnamefont {W.-K.}\ \bibnamefont {Mok}},
  \bibinfo {author} {\bibfnamefont {S.}~\bibnamefont {Sim}}, \bibinfo {author}
  {\bibfnamefont {L.-C.}\ \bibnamefont {Kwek}},\ and\ \bibinfo {author}
  {\bibfnamefont {A.}~\bibnamefont {Aspuru-Guzik}},\ }\href
  {https://doi.org/10.1103/RevModPhys.94.015004} {\bibfield  {journal}
  {\bibinfo  {journal} {Rev. Mod. Phys.}\ }\textbf {\bibinfo {volume} {94}},\
  \bibinfo {pages} {015004} (\bibinfo {year} {2022})}\BibitemShut {NoStop}%
\bibitem [{\citenamefont {Peruzzo}\ \emph {et~al.}(2014)\citenamefont
  {Peruzzo}, \citenamefont {McClean}, \citenamefont {Shadbolt}, \citenamefont
  {Yung}, \citenamefont {Zhou}, \citenamefont {Love}, \citenamefont
  {Aspuru-Guzik},\ and\ \citenamefont {O’brien}}]{peruzzo2014variational}%
  \BibitemOpen
  \bibfield  {author} {\bibinfo {author} {\bibfnamefont {A.}~\bibnamefont
  {Peruzzo}}, \bibinfo {author} {\bibfnamefont {J.}~\bibnamefont {McClean}},
  \bibinfo {author} {\bibfnamefont {P.}~\bibnamefont {Shadbolt}}, \bibinfo
  {author} {\bibfnamefont {M.-H.}\ \bibnamefont {Yung}}, \bibinfo {author}
  {\bibfnamefont {X.-Q.}\ \bibnamefont {Zhou}}, \bibinfo {author}
  {\bibfnamefont {P.~J.}\ \bibnamefont {Love}}, \bibinfo {author}
  {\bibfnamefont {A.}~\bibnamefont {Aspuru-Guzik}},\ and\ \bibinfo {author}
  {\bibfnamefont {J.~L.}\ \bibnamefont {O’brien}},\ }\href
  {https://doi.org/https://doi.org/10.1038/ncomms5213} {\bibfield  {journal}
  {\bibinfo  {journal} {Nat. Commun}\ }\textbf {\bibinfo {volume} {5}},\
  \bibinfo {pages} {4213} (\bibinfo {year} {2014})}\BibitemShut {NoStop}%
\bibitem [{\citenamefont {Cerezo}\ \emph {et~al.}(2021)\citenamefont {Cerezo},
  \citenamefont {Arrasmith}, \citenamefont {Babbush}, \citenamefont {Benjamin},
  \citenamefont {Endo}, \citenamefont {Fujii}, \citenamefont {McClean},
  \citenamefont {Mitarai}, \citenamefont {Yuan}, \citenamefont {Cincio},\ and\
  \citenamefont {Coles}}]{cerezo2021variational}%
  \BibitemOpen
  \bibfield  {author} {\bibinfo {author} {\bibfnamefont {M.}~\bibnamefont
  {Cerezo}}, \bibinfo {author} {\bibfnamefont {A.}~\bibnamefont {Arrasmith}},
  \bibinfo {author} {\bibfnamefont {R.}~\bibnamefont {Babbush}}, \bibinfo
  {author} {\bibfnamefont {S.~C.}\ \bibnamefont {Benjamin}}, \bibinfo {author}
  {\bibfnamefont {S.}~\bibnamefont {Endo}}, \bibinfo {author} {\bibfnamefont
  {K.}~\bibnamefont {Fujii}}, \bibinfo {author} {\bibfnamefont {J.~R.}\
  \bibnamefont {McClean}}, \bibinfo {author} {\bibfnamefont {K.}~\bibnamefont
  {Mitarai}}, \bibinfo {author} {\bibfnamefont {X.}~\bibnamefont {Yuan}},
  \bibinfo {author} {\bibfnamefont {L.}~\bibnamefont {Cincio}},\ and\ \bibinfo
  {author} {\bibfnamefont {P.~J.}\ \bibnamefont {Coles}},\ }\href
  {https://doi.org/https://doi.org/10.1038/s42254-021-00348-9} {\bibfield
  {journal} {\bibinfo  {journal} {Nat. Rev. Phys.}\ }\textbf {\bibinfo {volume}
  {3}},\ \bibinfo {pages} {625} (\bibinfo {year} {2021})}\BibitemShut {NoStop}%
\bibitem [{\citenamefont {Grimsley}\ \emph {et~al.}(2019)\citenamefont
  {Grimsley}, \citenamefont {Economou}, \citenamefont {Barnes},\ and\
  \citenamefont {Mayhall}}]{grimsley2019adaptive}%
  \BibitemOpen
  \bibfield  {author} {\bibinfo {author} {\bibfnamefont {H.~R.}\ \bibnamefont
  {Grimsley}}, \bibinfo {author} {\bibfnamefont {S.~E.}\ \bibnamefont
  {Economou}}, \bibinfo {author} {\bibfnamefont {E.}~\bibnamefont {Barnes}},\
  and\ \bibinfo {author} {\bibfnamefont {N.~J.}\ \bibnamefont {Mayhall}},\
  }\href {https://doi.org/https://doi.org/10.1038/s41467-019-10988-2}
  {\bibfield  {journal} {\bibinfo  {journal} {Nat. Commun}\ }\textbf {\bibinfo
  {volume} {10}},\ \bibinfo {pages} {3007} (\bibinfo {year}
  {2019})}\BibitemShut {NoStop}%
\bibitem [{\citenamefont {Gyawali}\ and\ \citenamefont
  {Lawler}(2022)}]{Gyawali2022adaptive}%
  \BibitemOpen
  \bibfield  {author} {\bibinfo {author} {\bibfnamefont {G.}~\bibnamefont
  {Gyawali}}\ and\ \bibinfo {author} {\bibfnamefont {M.~J.}\ \bibnamefont
  {Lawler}},\ }\href {https://doi.org/10.1103/PhysRevA.105.012413} {\bibfield
  {journal} {\bibinfo  {journal} {Phys. Rev. A}\ }\textbf {\bibinfo {volume}
  {105}},\ \bibinfo {pages} {012413} (\bibinfo {year} {2022})}\BibitemShut
  {NoStop}%
\bibitem [{\citenamefont {Burton}\ \emph {et~al.}(2023)\citenamefont {Burton},
  \citenamefont {Marti-Dafcik}, \citenamefont {Tew},\ and\ \citenamefont
  {Wales}}]{burton2023exact}%
  \BibitemOpen
  \bibfield  {author} {\bibinfo {author} {\bibfnamefont {H.~G.}\ \bibnamefont
  {Burton}}, \bibinfo {author} {\bibfnamefont {D.}~\bibnamefont
  {Marti-Dafcik}}, \bibinfo {author} {\bibfnamefont {D.~P.}\ \bibnamefont
  {Tew}},\ and\ \bibinfo {author} {\bibfnamefont {D.~J.}\ \bibnamefont
  {Wales}},\ }\href
  {https://doi.org/https://doi.org/10.1038/s41534-023-00744-2} {\bibfield
  {journal} {\bibinfo  {journal} {Npj Quantum Inf.}\ }\textbf {\bibinfo
  {volume} {9}},\ \bibinfo {pages} {75} (\bibinfo {year} {2023})}\BibitemShut
  {NoStop}%
\bibitem [{\citenamefont {Larsen}\ \emph {et~al.}(2024)\citenamefont {Larsen},
  \citenamefont {Grace}, \citenamefont {Baczewski},\ and\ \citenamefont
  {Magann}}]{larsen2024feedback}%
  \BibitemOpen
  \bibfield  {author} {\bibinfo {author} {\bibfnamefont {J.~B.}\ \bibnamefont
  {Larsen}}, \bibinfo {author} {\bibfnamefont {M.~D.}\ \bibnamefont {Grace}},
  \bibinfo {author} {\bibfnamefont {A.~D.}\ \bibnamefont {Baczewski}},\ and\
  \bibinfo {author} {\bibfnamefont {A.~B.}\ \bibnamefont {Magann}},\ }\href
  {https://doi.org/10.1103/PhysRevResearch.6.033336} {\bibfield  {journal}
  {\bibinfo  {journal} {Phys. Rev. Res.}\ }\textbf {\bibinfo {volume} {6}},\
  \bibinfo {pages} {033336} (\bibinfo {year} {2024})}\BibitemShut {NoStop}%
\bibitem [{\citenamefont {Magann}\ \emph {et~al.}(2022)\citenamefont {Magann},
  \citenamefont {Rudinger}, \citenamefont {Grace},\ and\ \citenamefont
  {Sarovar}}]{Magann2022feedback}%
  \BibitemOpen
  \bibfield  {author} {\bibinfo {author} {\bibfnamefont {A.~B.}\ \bibnamefont
  {Magann}}, \bibinfo {author} {\bibfnamefont {K.~M.}\ \bibnamefont
  {Rudinger}}, \bibinfo {author} {\bibfnamefont {M.~D.}\ \bibnamefont
  {Grace}},\ and\ \bibinfo {author} {\bibfnamefont {M.}~\bibnamefont
  {Sarovar}},\ }\href {https://doi.org/10.1103/PhysRevLett.129.250502}
  {\bibfield  {journal} {\bibinfo  {journal} {Phys. Rev. Lett.}\ }\textbf
  {\bibinfo {volume} {129}},\ \bibinfo {pages} {250502} (\bibinfo {year}
  {2022})}\BibitemShut {NoStop}%
\bibitem [{\citenamefont {Xie}\ \emph {et~al.}(2022)\citenamefont {Xie},
  \citenamefont {Seki},\ and\ \citenamefont {Yunoki}}]{xie2022variational}%
  \BibitemOpen
  \bibfield  {author} {\bibinfo {author} {\bibfnamefont {Q.}~\bibnamefont
  {Xie}}, \bibinfo {author} {\bibfnamefont {K.}~\bibnamefont {Seki}},\ and\
  \bibinfo {author} {\bibfnamefont {S.}~\bibnamefont {Yunoki}},\ }\href
  {https://doi.org/10.1103/PhysRevB.106.155153} {\bibfield  {journal} {\bibinfo
   {journal} {Phys. Rev. B}\ }\textbf {\bibinfo {volume} {106}},\ \bibinfo
  {pages} {155153} (\bibinfo {year} {2022})}\BibitemShut {NoStop}%
\bibitem [{\citenamefont {Claeys}\ \emph {et~al.}(2019)\citenamefont {Claeys},
  \citenamefont {Pandey}, \citenamefont {Sels},\ and\ \citenamefont
  {Polkovnikov}}]{claeys2019floquet}%
  \BibitemOpen
  \bibfield  {author} {\bibinfo {author} {\bibfnamefont {P.~W.}\ \bibnamefont
  {Claeys}}, \bibinfo {author} {\bibfnamefont {M.}~\bibnamefont {Pandey}},
  \bibinfo {author} {\bibfnamefont {D.}~\bibnamefont {Sels}},\ and\ \bibinfo
  {author} {\bibfnamefont {A.}~\bibnamefont {Polkovnikov}},\ }\href
  {https://doi.org/10.1103/PhysRevLett.123.090602} {\bibfield  {journal}
  {\bibinfo  {journal} {Phys. Rev. Lett.}\ }\textbf {\bibinfo {volume} {123}},\
  \bibinfo {pages} {090602} (\bibinfo {year} {2019})}\BibitemShut {NoStop}%
\bibitem [{\citenamefont {Tang}\ \emph {et~al.}(2024)\citenamefont {Tang},
  \citenamefont {Xu}, \citenamefont {Ding}, \citenamefont {Xu}, \citenamefont
  {Ban}, \citenamefont {Yung}, \citenamefont {P{\'e}rez-Obiol}, \citenamefont
  {Platero},\ and\ \citenamefont {Chen}}]{tang2024exploring}%
  \BibitemOpen
  \bibfield  {author} {\bibinfo {author} {\bibfnamefont {J.}~\bibnamefont
  {Tang}}, \bibinfo {author} {\bibfnamefont {R.}~\bibnamefont {Xu}}, \bibinfo
  {author} {\bibfnamefont {Y.}~\bibnamefont {Ding}}, \bibinfo {author}
  {\bibfnamefont {X.}~\bibnamefont {Xu}}, \bibinfo {author} {\bibfnamefont
  {Y.}~\bibnamefont {Ban}}, \bibinfo {author} {\bibfnamefont {M.-H.}\
  \bibnamefont {Yung}}, \bibinfo {author} {\bibfnamefont {A.}~\bibnamefont
  {P{\'e}rez-Obiol}}, \bibinfo {author} {\bibfnamefont {G.}~\bibnamefont
  {Platero}},\ and\ \bibinfo {author} {\bibfnamefont {X.}~\bibnamefont
  {Chen}},\ }\href {https://doi.org/https://doi.org/10.1038/s41535-024-00697-5}
  {\bibfield  {journal} {\bibinfo  {journal} {npj Quantum Mater.}\ }\textbf
  {\bibinfo {volume} {9}},\ \bibinfo {pages} {87} (\bibinfo {year}
  {2024})}\BibitemShut {NoStop}%
\bibitem [{\citenamefont {Marti}\ \emph {et~al.}(2024)\citenamefont {Marti},
  \citenamefont {Mansuroglu},\ and\ \citenamefont
  {Hartmann}}]{marti2024efficient}%
  \BibitemOpen
  \bibfield  {author} {\bibinfo {author} {\bibfnamefont {L.}~\bibnamefont
  {Marti}}, \bibinfo {author} {\bibfnamefont {R.}~\bibnamefont {Mansuroglu}},\
  and\ \bibinfo {author} {\bibfnamefont {M.~J.}\ \bibnamefont {Hartmann}},\
  }\bibfield  {journal} {\bibinfo  {journal} {arXiv preprint}\ }\href
  {https://doi.org/https://doi.org/10.48550/arXiv.2403.14506}
  {https://doi.org/10.48550/arXiv.2403.14506} (\bibinfo {year}
  {2024})\BibitemShut {NoStop}%
\bibitem [{\citenamefont {Reiner}\ \emph {et~al.}(2019)\citenamefont {Reiner},
  \citenamefont {Wilhelm-Mauch}, \citenamefont {Sch{\"o}n},\ and\ \citenamefont
  {Marthaler}}]{reiner2019finding}%
  \BibitemOpen
  \bibfield  {author} {\bibinfo {author} {\bibfnamefont {J.-M.}\ \bibnamefont
  {Reiner}}, \bibinfo {author} {\bibfnamefont {F.}~\bibnamefont
  {Wilhelm-Mauch}}, \bibinfo {author} {\bibfnamefont {G.}~\bibnamefont
  {Sch{\"o}n}},\ and\ \bibinfo {author} {\bibfnamefont {M.}~\bibnamefont
  {Marthaler}},\ }\href
  {https://doi.org/https://doi.org/10.1088/2058-9565/ab1e85} {\bibfield
  {journal} {\bibinfo  {journal} {QST}\ }\textbf {\bibinfo {volume} {4}},\
  \bibinfo {pages} {035005} (\bibinfo {year} {2019})}\BibitemShut {NoStop}%
\bibitem [{\citenamefont {Wecker}\ \emph {et~al.}(2015)\citenamefont {Wecker},
  \citenamefont {Hastings},\ and\ \citenamefont {Troyer}}]{wecker2015progress}%
  \BibitemOpen
  \bibfield  {author} {\bibinfo {author} {\bibfnamefont {D.}~\bibnamefont
  {Wecker}}, \bibinfo {author} {\bibfnamefont {M.~B.}\ \bibnamefont
  {Hastings}},\ and\ \bibinfo {author} {\bibfnamefont {M.}~\bibnamefont
  {Troyer}},\ }\href {https://doi.org/10.1103/PhysRevA.92.042303} {\bibfield
  {journal} {\bibinfo  {journal} {Phys. Rev. A}\ }\textbf {\bibinfo {volume}
  {92}},\ \bibinfo {pages} {042303} (\bibinfo {year} {2015})}\BibitemShut
  {NoStop}%
\bibitem [{\citenamefont {Cade}\ \emph {et~al.}(2020)\citenamefont {Cade},
  \citenamefont {Mineh}, \citenamefont {Montanaro},\ and\ \citenamefont
  {Stanisic}}]{Cade2020Strategy}%
  \BibitemOpen
  \bibfield  {author} {\bibinfo {author} {\bibfnamefont {C.}~\bibnamefont
  {Cade}}, \bibinfo {author} {\bibfnamefont {L.}~\bibnamefont {Mineh}},
  \bibinfo {author} {\bibfnamefont {A.}~\bibnamefont {Montanaro}},\ and\
  \bibinfo {author} {\bibfnamefont {S.}~\bibnamefont {Stanisic}},\ }\href
  {https://doi.org/10.1103/PhysRevB.102.235122} {\bibfield  {journal} {\bibinfo
   {journal} {Phys. Rev. B}\ }\textbf {\bibinfo {volume} {102}},\ \bibinfo
  {pages} {235122} (\bibinfo {year} {2020})}\BibitemShut {NoStop}%
\bibitem [{\citenamefont {Cai}(2020)}]{cai2020resource}%
  \BibitemOpen
  \bibfield  {author} {\bibinfo {author} {\bibfnamefont {Z.}~\bibnamefont
  {Cai}},\ }\href {https://doi.org/10.1103/PhysRevApplied.14.014059} {\bibfield
   {journal} {\bibinfo  {journal} {Phys. Rev. Appl.}\ }\textbf {\bibinfo
  {volume} {14}},\ \bibinfo {pages} {014059} (\bibinfo {year}
  {2020})}\BibitemShut {NoStop}%
\bibitem [{\citenamefont {Alvertis}\ \emph {et~al.}(2025)\citenamefont
  {Alvertis}, \citenamefont {Khan}, \citenamefont {Iadecola}, \citenamefont
  {Orth},\ and\ \citenamefont {Tubman}}]{alvertis2025classical}%
  \BibitemOpen
  \bibfield  {author} {\bibinfo {author} {\bibfnamefont {A.~M.}\ \bibnamefont
  {Alvertis}}, \bibinfo {author} {\bibfnamefont {A.}~\bibnamefont {Khan}},
  \bibinfo {author} {\bibfnamefont {T.}~\bibnamefont {Iadecola}}, \bibinfo
  {author} {\bibfnamefont {P.~P.}\ \bibnamefont {Orth}},\ and\ \bibinfo
  {author} {\bibfnamefont {N.}~\bibnamefont {Tubman}},\ }\href
  {https://doi.org/https://doi.org/10.22331/q-2025-05-20-1748} {\bibfield
  {journal} {\bibinfo  {journal} {Quantum}\ }\textbf {\bibinfo {volume} {9}},\
  \bibinfo {pages} {1748} (\bibinfo {year} {2025})}\BibitemShut {NoStop}%
\bibitem [{\citenamefont {Breuer}\ and\ \citenamefont
  {Petruccione}(2002)}]{breuer2002theory}%
  \BibitemOpen
  \bibfield  {author} {\bibinfo {author} {\bibfnamefont {H.-P.}\ \bibnamefont
  {Breuer}}\ and\ \bibinfo {author} {\bibfnamefont {F.}~\bibnamefont
  {Petruccione}},\ }\href
  {https://doi.org/https://doi.org/10.1093/acprof:oso/9780199213900.001.0001,}
  {\emph {\bibinfo {title} {The theory of open quantum systems}}}\ (\bibinfo
  {publisher} {Oxford University Press, USA},\ \bibinfo {year}
  {2002})\BibitemShut {NoStop}%
\bibitem [{\citenamefont {Izmaylov}\ \emph {et~al.}(2020)\citenamefont
  {Izmaylov}, \citenamefont {D{\'\i}az-Tinoco},\ and\ \citenamefont
  {Lang}}]{izmaylov2020order}%
  \BibitemOpen
  \bibfield  {author} {\bibinfo {author} {\bibfnamefont {A.~F.}\ \bibnamefont
  {Izmaylov}}, \bibinfo {author} {\bibfnamefont {M.}~\bibnamefont
  {D{\'\i}az-Tinoco}},\ and\ \bibinfo {author} {\bibfnamefont {R.~A.}\
  \bibnamefont {Lang}},\ }\href
  {https://doi.org/https://doi.org/10.1039/D0CP01707H} {\bibfield  {journal}
  {\bibinfo  {journal} {Phys. Chem. Chem.}\ }\textbf {\bibinfo {volume} {22}},\
  \bibinfo {pages} {12980} (\bibinfo {year} {2020})}\BibitemShut {NoStop}%
\bibitem [{\citenamefont {Jordan}\ and\ \citenamefont
  {Wigner}(1928)}]{jordan1993paulische}%
  \BibitemOpen
  \bibfield  {author} {\bibinfo {author} {\bibfnamefont {P.}~\bibnamefont
  {Jordan}}\ and\ \bibinfo {author} {\bibfnamefont {E.}~\bibnamefont
  {Wigner}},\ }\href@noop {} {\bibfield  {journal} {\bibinfo  {journal}
  {Zeitschrift für Physik}\ }\textbf {\bibinfo {volume} {47}},\ \bibinfo
  {pages} {631} (\bibinfo {year} {1928})}\BibitemShut {NoStop}%
\bibitem [{\citenamefont {K{\"o}kc{\"u}}\ \emph {et~al.}(2022)\citenamefont
  {K{\"o}kc{\"u}}, \citenamefont {Camps}, \citenamefont {Bassman~Oftelie},
  \citenamefont {Freericks}, \citenamefont {de~Jong}, \citenamefont
  {Van~Beeumen},\ and\ \citenamefont {Kemper}}]{kokcu2022algebraic}%
  \BibitemOpen
  \bibfield  {author} {\bibinfo {author} {\bibfnamefont {E.}~\bibnamefont
  {K{\"o}kc{\"u}}}, \bibinfo {author} {\bibfnamefont {D.}~\bibnamefont
  {Camps}}, \bibinfo {author} {\bibfnamefont {L.}~\bibnamefont
  {Bassman~Oftelie}}, \bibinfo {author} {\bibfnamefont {J.~K.}\ \bibnamefont
  {Freericks}}, \bibinfo {author} {\bibfnamefont {W.~A.}\ \bibnamefont
  {de~Jong}}, \bibinfo {author} {\bibfnamefont {R.}~\bibnamefont
  {Van~Beeumen}},\ and\ \bibinfo {author} {\bibfnamefont {A.~F.}\ \bibnamefont
  {Kemper}},\ }\href
  {https://doi.org/https://doi.org/10.1103/PhysRevA.105.032420} {\bibfield
  {journal} {\bibinfo  {journal} {Phys. Rev. A.}\ }\textbf {\bibinfo {volume}
  {105}},\ \bibinfo {pages} {032420} (\bibinfo {year} {2022})}\BibitemShut
  {NoStop}%
\bibitem [{\citenamefont {K{\"o}kc{\"u}}\ \emph {et~al.}(2023)\citenamefont
  {K{\"o}kc{\"u}}, \citenamefont {Camps}, \citenamefont {Oftelie},
  \citenamefont {de~Jong}, \citenamefont {Beeumen},\ and\ \citenamefont
  {Kemper}}]{kokcu2023algebraic}%
  \BibitemOpen
  \bibfield  {author} {\bibinfo {author} {\bibfnamefont {E.}~\bibnamefont
  {K{\"o}kc{\"u}}}, \bibinfo {author} {\bibfnamefont {D.}~\bibnamefont
  {Camps}}, \bibinfo {author} {\bibfnamefont {L.~B.}\ \bibnamefont {Oftelie}},
  \bibinfo {author} {\bibfnamefont {W.~A.}\ \bibnamefont {de~Jong}}, \bibinfo
  {author} {\bibfnamefont {R.~V.}\ \bibnamefont {Beeumen}},\ and\ \bibinfo
  {author} {\bibfnamefont {A.~F.}\ \bibnamefont {Kemper}},\ }\bibfield
  {journal} {\bibinfo  {journal} {arXiv preprint}\ }\href
  {https://doi.org/arXiv.2303.09538} {arXiv.2303.09538} (\bibinfo {year}
  {2023})\BibitemShut {NoStop}%
\bibitem [{\citenamefont {Javadi-Abhari}\ \emph {et~al.}(2024)\citenamefont
  {Javadi-Abhari}, \citenamefont {Treinish}, \citenamefont {Krsulich},
  \citenamefont {Wood}, \citenamefont {Lishman}, \citenamefont {Gacon},
  \citenamefont {Martiel}, \citenamefont {Nation}, \citenamefont {Bishop},
  \citenamefont {Cross}, \citenamefont {Johnson},\ and\ \citenamefont
  {Gambetta}}]{qiskit2024}%
  \BibitemOpen
  \bibfield  {author} {\bibinfo {author} {\bibfnamefont {A.}~\bibnamefont
  {Javadi-Abhari}}, \bibinfo {author} {\bibfnamefont {M.}~\bibnamefont
  {Treinish}}, \bibinfo {author} {\bibfnamefont {K.}~\bibnamefont {Krsulich}},
  \bibinfo {author} {\bibfnamefont {C.~J.}\ \bibnamefont {Wood}}, \bibinfo
  {author} {\bibfnamefont {J.}~\bibnamefont {Lishman}}, \bibinfo {author}
  {\bibfnamefont {J.}~\bibnamefont {Gacon}}, \bibinfo {author} {\bibfnamefont
  {S.}~\bibnamefont {Martiel}}, \bibinfo {author} {\bibfnamefont {P.~D.}\
  \bibnamefont {Nation}}, \bibinfo {author} {\bibfnamefont {L.~S.}\
  \bibnamefont {Bishop}}, \bibinfo {author} {\bibfnamefont {A.~W.}\
  \bibnamefont {Cross}}, \bibinfo {author} {\bibfnamefont {B.~R.}\ \bibnamefont
  {Johnson}},\ and\ \bibinfo {author} {\bibfnamefont {J.~M.}\ \bibnamefont
  {Gambetta}},\ }\href {https://doi.org/10.48550/arXiv.2405.08810} {\bibinfo
  {title} {Quantum computing with {Q}iskit}} (\bibinfo {year} {2024}),\ \Eprint
  {https://arxiv.org/abs/2405.08810} {arXiv:2405.08810 [quant-ph]} \BibitemShut
  {NoStop}%
\bibitem [{\citenamefont {Developers}(2025)}]{Cirq_Developers_2025}%
  \BibitemOpen
  \bibfield  {author} {\bibinfo {author} {\bibfnamefont {C.}~\bibnamefont
  {Developers}},\ }\href {https://doi.org/10.5281/ZENODO.4062499} {\emph
  {\bibinfo {title} {Cirq}}}\ (\bibinfo  {publisher} {Zenodo},\ \bibinfo {year}
  {2025})\BibitemShut {NoStop}%
\bibitem [{\citenamefont {B{\"a}umer}\ \emph {et~al.}(2024)\citenamefont
  {B{\"a}umer}, \citenamefont {Tripathi}, \citenamefont {Wang}, \citenamefont
  {Rall}, \citenamefont {Chen}, \citenamefont {Majumder}, \citenamefont
  {Seif},\ and\ \citenamefont {Minev}}]{baumer2024efficient}%
  \BibitemOpen
  \bibfield  {author} {\bibinfo {author} {\bibfnamefont {E.}~\bibnamefont
  {B{\"a}umer}}, \bibinfo {author} {\bibfnamefont {V.}~\bibnamefont
  {Tripathi}}, \bibinfo {author} {\bibfnamefont {D.~S.}\ \bibnamefont {Wang}},
  \bibinfo {author} {\bibfnamefont {P.}~\bibnamefont {Rall}}, \bibinfo {author}
  {\bibfnamefont {E.~H.}\ \bibnamefont {Chen}}, \bibinfo {author}
  {\bibfnamefont {S.}~\bibnamefont {Majumder}}, \bibinfo {author}
  {\bibfnamefont {A.}~\bibnamefont {Seif}},\ and\ \bibinfo {author}
  {\bibfnamefont {Z.~K.}\ \bibnamefont {Minev}},\ }\href
  {https://doi.org/https://doi.org/10.1103/PRXQuantum.5.030339} {\bibfield
  {journal} {\bibinfo  {journal} {PRX Quantum}\ }\textbf {\bibinfo {volume}
  {5}},\ \bibinfo {pages} {030339} (\bibinfo {year} {2024})}\BibitemShut
  {NoStop}%
\bibitem [{\citenamefont {Acharya}\ \emph {et~al.}(2024)\citenamefont
  {Acharya}, \citenamefont {Abanin}, \citenamefont {Aghababaie-Beni},
  \citenamefont {Aleiner}, \citenamefont {Andersen}, \citenamefont {Ansmann},
  \citenamefont {Arute}, \citenamefont {Arya}, \citenamefont {Asfaw},
  \citenamefont {Astrakhantsev} \emph {et~al.}}]{acharya2024quantum}%
  \BibitemOpen
  \bibfield  {author} {\bibinfo {author} {\bibfnamefont {R.}~\bibnamefont
  {Acharya}}, \bibinfo {author} {\bibfnamefont {D.~A.}\ \bibnamefont {Abanin}},
  \bibinfo {author} {\bibfnamefont {L.}~\bibnamefont {Aghababaie-Beni}},
  \bibinfo {author} {\bibfnamefont {I.}~\bibnamefont {Aleiner}}, \bibinfo
  {author} {\bibfnamefont {T.~I.}\ \bibnamefont {Andersen}}, \bibinfo {author}
  {\bibfnamefont {M.}~\bibnamefont {Ansmann}}, \bibinfo {author} {\bibfnamefont
  {F.}~\bibnamefont {Arute}}, \bibinfo {author} {\bibfnamefont
  {K.}~\bibnamefont {Arya}}, \bibinfo {author} {\bibfnamefont {A.}~\bibnamefont
  {Asfaw}}, \bibinfo {author} {\bibfnamefont {N.}~\bibnamefont
  {Astrakhantsev}}, \emph {et~al.},\ }\bibfield  {journal} {\bibinfo  {journal}
  {Nature}\ }\href {https://doi.org/https://doi.org/10.1038/s41586-024-08449-y}
  {https://doi.org/10.1038/s41586-024-08449-y} (\bibinfo {year}
  {2024})\BibitemShut {NoStop}%
\bibitem [{\citenamefont {Betts}\ \emph {et~al.}(1996)\citenamefont {Betts},
  \citenamefont {Masui}, \citenamefont {Vats},\ and\ \citenamefont
  {Stewart}}]{betts1996improved}%
  \BibitemOpen
  \bibfield  {author} {\bibinfo {author} {\bibfnamefont {D.}~\bibnamefont
  {Betts}}, \bibinfo {author} {\bibfnamefont {S.}~\bibnamefont {Masui}},
  \bibinfo {author} {\bibfnamefont {N.}~\bibnamefont {Vats}},\ and\ \bibinfo
  {author} {\bibfnamefont {G.}~\bibnamefont {Stewart}},\ }\href
  {https://doi.org/https://doi.org/10.1139/p96-010} {\bibfield  {journal}
  {\bibinfo  {journal} {Can. J. Phys.}\ }\textbf {\bibinfo {volume} {74}},\
  \bibinfo {pages} {54} (\bibinfo {year} {1996})}\BibitemShut {NoStop}%
\bibitem [{\citenamefont {Noack}\ \emph {et~al.}(1996)\citenamefont {Noack},
  \citenamefont {White},\ and\ \citenamefont {Scalapino}}]{noack1996ground}%
  \BibitemOpen
  \bibfield  {author} {\bibinfo {author} {\bibfnamefont {R.}~\bibnamefont
  {Noack}}, \bibinfo {author} {\bibfnamefont {S.}~\bibnamefont {White}},\ and\
  \bibinfo {author} {\bibfnamefont {D.}~\bibnamefont {Scalapino}},\ }\href
  {https://doi.org/https://doi.org/10.1016/S0921-4534(96)00515-1} {\bibfield
  {journal} {\bibinfo  {journal} {Physica C: Superconductivity}\ }\textbf
  {\bibinfo {volume} {270}},\ \bibinfo {pages} {281} (\bibinfo {year}
  {1996})}\BibitemShut {NoStop}%
\bibitem [{\citenamefont {Kivlichan}\ \emph {et~al.}(2018)\citenamefont
  {Kivlichan}, \citenamefont {McClean}, \citenamefont {Wiebe}, \citenamefont
  {Gidney}, \citenamefont {Aspuru-Guzik}, \citenamefont {Chan},\ and\
  \citenamefont {Babbush}}]{kivlichan2018quantum}%
  \BibitemOpen
  \bibfield  {author} {\bibinfo {author} {\bibfnamefont {I.~D.}\ \bibnamefont
  {Kivlichan}}, \bibinfo {author} {\bibfnamefont {J.}~\bibnamefont {McClean}},
  \bibinfo {author} {\bibfnamefont {N.}~\bibnamefont {Wiebe}}, \bibinfo
  {author} {\bibfnamefont {C.}~\bibnamefont {Gidney}}, \bibinfo {author}
  {\bibfnamefont {A.}~\bibnamefont {Aspuru-Guzik}}, \bibinfo {author}
  {\bibfnamefont {G.~K.-L.}\ \bibnamefont {Chan}},\ and\ \bibinfo {author}
  {\bibfnamefont {R.}~\bibnamefont {Babbush}},\ }\href
  {https://doi.org/https://doi.org/10.1103/PhysRevLett.120.110501} {\bibfield
  {journal} {\bibinfo  {journal} {Phys. Rev. Lett.}\ }\textbf {\bibinfo
  {volume} {120}},\ \bibinfo {pages} {110501} (\bibinfo {year}
  {2018})}\BibitemShut {NoStop}%
\bibitem [{\citenamefont {Gros}(1996)}]{gros1996control}%
  \BibitemOpen
  \bibfield  {author} {\bibinfo {author} {\bibfnamefont {C.}~\bibnamefont
  {Gros}},\ }\href {https://doi.org/10.1103/PhysRevB.53.6865} {\bibfield
  {journal} {\bibinfo  {journal} {Phys. Rev. B}\ }\textbf {\bibinfo {volume}
  {53}},\ \bibinfo {pages} {6865} (\bibinfo {year} {1996})}\BibitemShut
  {NoStop}%
\bibitem [{\citenamefont {Jiang}\ \emph {et~al.}(2018)\citenamefont {Jiang},
  \citenamefont {Sung}, \citenamefont {Kechedzhi}, \citenamefont
  {Smelyanskiy},\ and\ \citenamefont {Boixo}}]{Jiang2018quantum}%
  \BibitemOpen
  \bibfield  {author} {\bibinfo {author} {\bibfnamefont {Z.}~\bibnamefont
  {Jiang}}, \bibinfo {author} {\bibfnamefont {K.~J.}\ \bibnamefont {Sung}},
  \bibinfo {author} {\bibfnamefont {K.}~\bibnamefont {Kechedzhi}}, \bibinfo
  {author} {\bibfnamefont {V.~N.}\ \bibnamefont {Smelyanskiy}},\ and\ \bibinfo
  {author} {\bibfnamefont {S.}~\bibnamefont {Boixo}},\ }\href
  {https://doi.org/10.1103/PhysRevApplied.9.044036} {\bibfield  {journal}
  {\bibinfo  {journal} {Phys. Rev. Appl.}\ }\textbf {\bibinfo {volume} {9}},\
  \bibinfo {pages} {044036} (\bibinfo {year} {2018})}\BibitemShut {NoStop}%
\bibitem [{\citenamefont {Verstraete}\ \emph {et~al.}(2009)\citenamefont
  {Verstraete}, \citenamefont {Cirac},\ and\ \citenamefont
  {Latorre}}]{Verstraete2009quantum}%
  \BibitemOpen
  \bibfield  {author} {\bibinfo {author} {\bibfnamefont {F.}~\bibnamefont
  {Verstraete}}, \bibinfo {author} {\bibfnamefont {J.~I.}\ \bibnamefont
  {Cirac}},\ and\ \bibinfo {author} {\bibfnamefont {J.~I.}\ \bibnamefont
  {Latorre}},\ }\href {https://doi.org/10.1103/PhysRevA.79.032316} {\bibfield
  {journal} {\bibinfo  {journal} {Phys. Rev. A}\ }\textbf {\bibinfo {volume}
  {79}},\ \bibinfo {pages} {032316} (\bibinfo {year} {2009})}\BibitemShut
  {NoStop}%
\bibitem [{\citenamefont {Z.He}(2025)}]{data}%
  \BibitemOpen
  \bibfield  {author} {\bibinfo {author} {\bibfnamefont {J.~K.~F.}\
  \bibnamefont {Z.He}, \bibfnamefont {A.~F.~Kemper}},\ }\href
  {https://doi.org/10.5281/zenodo.14722297} {\bibinfo {title} {Dataset}}
  (\bibinfo {year} {2025})\BibitemShut {NoStop}%
\bibitem [{\citenamefont {Paszke}\ \emph {et~al.}(2017)\citenamefont {Paszke},
  \citenamefont {Gross}, \citenamefont {Chintala}, \citenamefont {Chanan},
  \citenamefont {Yang}, \citenamefont {DeVito}, \citenamefont {Lin},
  \citenamefont {Desmaison}, \citenamefont {Antiga},\ and\ \citenamefont
  {Lerer}}]{paszke2017automatic}%
  \BibitemOpen
  \bibfield  {author} {\bibinfo {author} {\bibfnamefont {A.}~\bibnamefont
  {Paszke}}, \bibinfo {author} {\bibfnamefont {S.}~\bibnamefont {Gross}},
  \bibinfo {author} {\bibfnamefont {S.}~\bibnamefont {Chintala}}, \bibinfo
  {author} {\bibfnamefont {G.}~\bibnamefont {Chanan}}, \bibinfo {author}
  {\bibfnamefont {E.}~\bibnamefont {Yang}}, \bibinfo {author} {\bibfnamefont
  {Z.}~\bibnamefont {DeVito}}, \bibinfo {author} {\bibfnamefont
  {Z.}~\bibnamefont {Lin}}, \bibinfo {author} {\bibfnamefont {A.}~\bibnamefont
  {Desmaison}}, \bibinfo {author} {\bibfnamefont {L.}~\bibnamefont {Antiga}},\
  and\ \bibinfo {author} {\bibfnamefont {A.}~\bibnamefont {Lerer}},\
  }\href@noop {} {\  (\bibinfo {year} {2017})}\BibitemShut {NoStop}%
\bibitem [{\citenamefont {Stanisic}\ \emph {et~al.}(2022)\citenamefont
  {Stanisic}, \citenamefont {Bosse}, \citenamefont {Gambetta}, \citenamefont
  {Santos}, \citenamefont {Mruczkiewicz}, \citenamefont {O’Brien},
  \citenamefont {Ostby},\ and\ \citenamefont
  {Montanaro}}]{stanisic2022observing}%
  \BibitemOpen
  \bibfield  {author} {\bibinfo {author} {\bibfnamefont {S.}~\bibnamefont
  {Stanisic}}, \bibinfo {author} {\bibfnamefont {J.~L.}\ \bibnamefont {Bosse}},
  \bibinfo {author} {\bibfnamefont {F.~M.}\ \bibnamefont {Gambetta}}, \bibinfo
  {author} {\bibfnamefont {R.~A.}\ \bibnamefont {Santos}}, \bibinfo {author}
  {\bibfnamefont {W.}~\bibnamefont {Mruczkiewicz}}, \bibinfo {author}
  {\bibfnamefont {T.~E.}\ \bibnamefont {O’Brien}}, \bibinfo {author}
  {\bibfnamefont {E.}~\bibnamefont {Ostby}},\ and\ \bibinfo {author}
  {\bibfnamefont {A.}~\bibnamefont {Montanaro}},\ }\href
  {https://doi.org/https://doi.org/10.1038/s41467-022-33335-4} {\bibfield
  {journal} {\bibinfo  {journal} {Nat. Commun.}\ }\textbf {\bibinfo {volume}
  {13}},\ \bibinfo {pages} {5743} (\bibinfo {year} {2022})}\BibitemShut
  {NoStop}%
\bibitem [{\citenamefont {Motta}\ \emph {et~al.}(2023)\citenamefont {Motta},
  \citenamefont {Sung}, \citenamefont {Whaley}, \citenamefont {Head-Gordon},\
  and\ \citenamefont {Shee}}]{motta2023bridging}%
  \BibitemOpen
  \bibfield  {author} {\bibinfo {author} {\bibfnamefont {M.}~\bibnamefont
  {Motta}}, \bibinfo {author} {\bibfnamefont {K.~J.}\ \bibnamefont {Sung}},
  \bibinfo {author} {\bibfnamefont {K.~B.}\ \bibnamefont {Whaley}}, \bibinfo
  {author} {\bibfnamefont {M.}~\bibnamefont {Head-Gordon}},\ and\ \bibinfo
  {author} {\bibfnamefont {J.}~\bibnamefont {Shee}},\ }\href
  {https://doi.org/https://doi.org/10.1039/D3SC02516K} {\bibfield  {journal}
  {\bibinfo  {journal} {Chem. Sci.}\ }\textbf {\bibinfo {volume} {14}},\
  \bibinfo {pages} {11213} (\bibinfo {year} {2023})}\BibitemShut {NoStop}%
\bibitem [{\citenamefont {Chen}\ \emph {et~al.}(2021)\citenamefont {Chen},
  \citenamefont {Cheng},\ and\ \citenamefont {Freericks}}]{chen2021quantum}%
  \BibitemOpen
  \bibfield  {author} {\bibinfo {author} {\bibfnamefont {J.}~\bibnamefont
  {Chen}}, \bibinfo {author} {\bibfnamefont {H.-P.}\ \bibnamefont {Cheng}},\
  and\ \bibinfo {author} {\bibfnamefont {J.~K.}\ \bibnamefont {Freericks}},\
  }\href {https://doi.org/https://doi.org/10.1021/acs.jctc.0c01052} {\bibfield
  {journal} {\bibinfo  {journal} {J. Chem. Theory Comput.}\ }\textbf {\bibinfo
  {volume} {17}},\ \bibinfo {pages} {841} (\bibinfo {year} {2021})}\BibitemShut
  {NoStop}%
\end{thebibliography}%

\appendix

\renewcommand{\theequation}{A\arabic{equation}}
\setcounter{equation}{0}

\renewcommand{\thefigure}{A\arabic{figure}}
\setcounter{figure}{0}
\renewcommand{\thefigure}{A\arabic{figure}}
\renewcommand{\theHfigure}{A\arabic{figure}}

~\\
\section{Initial state examples }
\label{sec:Initial state examples}
The only requirement for the initial state is to have the target value of total spin. But, we find it is advantageous to intersperse the doubly occupied sites and the empty sites, so that the hopping terms in the ansatz can reduce the double occupancy in the first layer. Hence, we select the simplest states that fulfill this requirement, which is a product state. We illustrate this with an eight-site example.

For the \( S = 0 \) sector, the choice is as follows:
\begin{equation}
\left|\psi_0\right\rangle = | \overbrace{\uparrow \downarrow }^{\text{site } 1} 
\underbrace{00}_{\text{site 2}} 
\overbrace{\uparrow \downarrow }^{\text{site } 3} 
\underbrace{00}_{\text{site 4}} 
\overbrace{\uparrow \downarrow }^{\text{site } 5} 
\underbrace{00}_{\text{site 6}} 
\overbrace{\uparrow \downarrow }^{\text{site } 7} 
\underbrace{00}_{\text{site 8}} \rangle
\end{equation}

For nonzero \( S \) sector initial states, we work with states that have \( S_z = S \). We assign double occupancies to \( \tfrac{N}{2} - S \) lattice sites, with the remaining electrons being spin-up on \( 2S \) sites, leaving $\tfrac{N}{2}-S$ empty sites. For example, for the \( S = 1 \) sector, a possible starting state is
\begin{equation}
\left|\psi_0\right\rangle = | \overbrace{00 }^{\text{site } 1} 
\underbrace{00}_{\text{site 2}} 
\overbrace{00 }^{\text{site } 3} 
\underbrace{\uparrow}_{\text{site 4}} 
\overbrace{\uparrow }^{\text{site } 5} 
\underbrace{\uparrow \downarrow}_{\text{site 6}} 
\overbrace{\uparrow \downarrow }^{\text{site } 7} 
\underbrace{\uparrow \downarrow}_{\text{site 8}} \rangle
\end{equation}

\renewcommand{\theequation}{B\arabic{equation}}
\setcounter{equation}{0}

\renewcommand{\thefigure}{B\arabic{figure}}
\setcounter{figure}{0}
\renewcommand{\thefigure}{B\arabic{figure}}
\renewcommand{\theHfigure}{B\arabic{figure}}

\section{Numerical optimization details}
\label{sec:Numerical optimization details}
\begin{figure*}
    \begin{centering}
        \includegraphics[width=2\columnwidth]{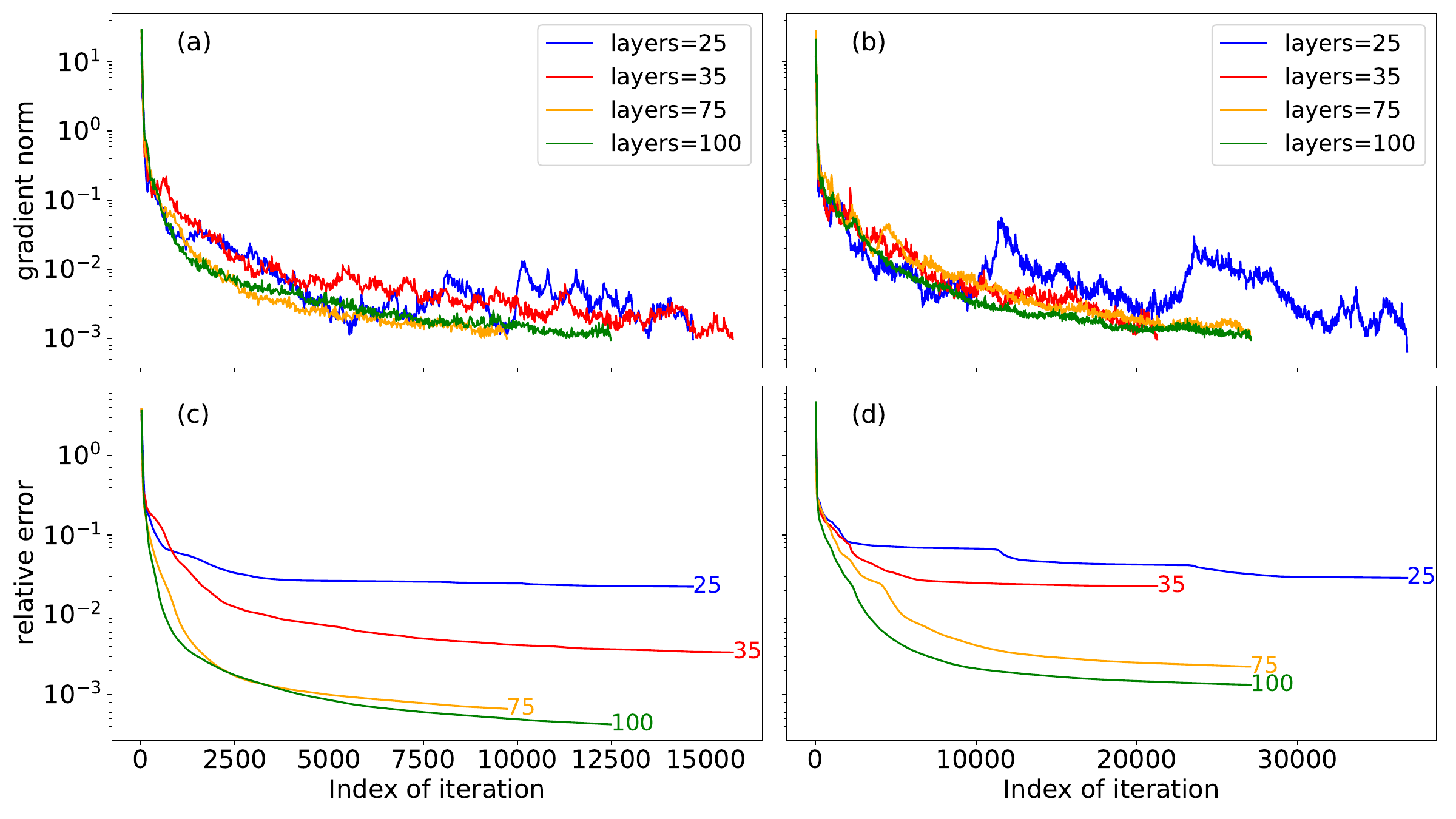}
        \caption{Gradient norms in panels (a) and (b) and relative errors in panels (c) and (d) during optimization over different numbers of layers. The \( x \)-axis represents the iteration index. The left panels, (a) and (c), correspond to the \( 2 \times 4 \) system, while the right panels, (b) and (d), correspond to the \( 2 \times 5 \) system.}
        \label{fig:numerical_peroformance}
    \end{centering}
\end{figure*}

We use the L-BFGS algorithm from the GPU-supported PyTorch package for the optimization~\cite{paszke2017automatic}. The choice of this optimizer is empirical, as L-BFGS, a second-order gradient method using an approximate Hessian matrix, often outperforms first-order optimizers such as ADAM. This preference has also been observed in many other variational studies~\cite{Cade2020Strategy,grimsley2019adaptive,stanisic2022observing,motta2023bridging}.

The parameter amplitudes are initialized as random values within the range \([-0.001, 0.001]\). By choosing a small random initialization, we ensure that the final optimized energy exhibits only a small numerical variation. The precise variance in the final energy due to the initial guess is case-dependent, as factors such as the potential parameter \(U\) and the number of sites can contribute to this variation. In this work, using the mentioned initial guess, we find that the variation in the final energy is typically at most on the order of \(10^{-2}\).

Though the goal of this work is to implement this algorithm on a quantum computer, we are currently using classical computer simulations. For the exponential of the hopping terms, we use an SU(2) identity~\cite{chen2021quantum} to implement the hopping terms instead of directly calculating the matrix exponential, which speeds up the computation. This is given by
\begin{equation}
\begin{split}
\exp \left[ i\theta\left( \hat{c}_{i}^\dagger \hat{c}_{a} + \hat{c}_{a}^\dagger \hat{c}_{i} \right) \right]
= \; & \mathbb{I} + i \sin \theta \left( \hat{c}_{i}^\dagger \hat{c}_{a} + \hat{c}_{a}^\dagger \hat{c}_{i} \right) \\
& + (\cos \theta - 1) \left( \hat{n}_{a} + \hat{n}_{i} \right) \\
& - 2 (\cos \theta - 1) \hat{n}_{a} \hat{n}_{i}.
\end{split}
\label{eq:su2_1}
\end{equation}
The stopping threshold for the optimization in this work is set to a gradient \( L_2 \) norm smaller than 0.001.

In Fig.~\ref{fig:numerical_peroformance}, we provide details of the convergence behavior for the two-leg ladder system with periodic boundary conditions, along with the number of iterations required as the system size increases. We present results for \( U/t = 8 \) only, as the total spin value of the ground state differs in the \( U/t = 2 \) regime for the \( 2 \times 4 \) and \( 2 \times 5 \) systems, which can affect performance. The results show that the optimization typically reduces the relative error to below 0.01 within a few thousand iterations for both system sizes. However, achieving the next order of accuracy, 0.001, often requires 5–7 times more iterations, as both the gradient and the step size along the gradient direction become very small when the optimization approaches the local minimum.

\end{document}